\documentclass{elsart1p} 
\usepackage{epsfig} 
\usepackage{amssymb} 
\usepackage[usenames,dvipsnames]{color} 

\journal{Nuclear Physics A} 

\begin{document} 

\begin{frontmatter} 

\title{First observation of the hyper superheavy hydrogen $^{6}_{\Lambda}$H} 
%\author[a]{The FINUDA Collaboration\thanksref{cor1}} 
%\linebreak 
%and 
%\linebreak 
%\author[b]{A.~Gal}, 
%\linebreak 
%\address[a]{INFN, Italy} 
%\address[b]{Racah Institute of Physics, 
%The Hebrew University, Jerusalem 91904, Israel} 
%\thanks[cor1]{Corresponding author: E. Botta, 
%e-mail address: botta@to.infn.it} 
\author[a,b]{M.~Agnello}, \author[c]{L.~Benussi}, \author[c]{M.~Bertani}, 
\author[d]{H.C.~Bhang}, \author[e,f]{G.~Bonomi}, 
\author[g,b]{E.~Botta\thanksref{cor1}}, \author[h,i]{M.~Bregant}, 
\author[g,b]{T.~Bressani}, \author[b]{S.~Bufalino}, \author[g,b]{L.~Busso}, 
\author[b]{D.~Calvo}, \author[h,i]{P.~Camerini}, \author[z]{B.~Dalena}, 
\author[g,b]{F.~De Mori}, \author[k,l]{G.~D'Erasmo}, \author[c]{F.L.~Fabbri}, 
\author[b]{A.~Feliciello}, \author[b]{A.~Filippi}, \author[k,l]{E.M.~Fiore}, 
\author[f]{A.~Fontana}, \author[m]{H.~Fujioka}, \author[f]{P.~Genova}, 
\author[c]{P.~Gianotti}, \author[i]{N.~Grion}, \author[c]{V.~Lucherini}, 
\author[g,b]{S.~Marcello}, \author[n]{N.~Mirfakhrai}, \author[e,f]{F.~Moia}, 
\author[o,b]{O.~Morra}, \author[m]{T.~Nagae}, \author[p]{H.~Outa}, 
\author[l]{A.~Pantaleo\thanksref{cor2}}, \author[l]{V.~Paticchio}, 
\author[i]{S.~Piano}, \author[h,i]{R.~Rui}, \author[k,l]{G.~Simonetti}, 
\author[b]{R.~Wheadon}, \author[e,f]{A.~Zenoni} 
\linebreak 
\author{(The FINUDA Collaboration)} 
\linebreak 
and 
\linebreak 
\author[q]{A.~Gal} 
\linebreak 

\address[a]{Dipartimento di Fisica, Politecnico di Torino, Corso Duca degli
Abruzzi 24, Torino, Italy} 
\address[b]{INFN Sezione di Torino, via P. Giuria 1, Torino, Italy} 
%\address[c]{INFN Sezione di Bari, via Amendola 173, Bari, Italy} 
\address[c]{Laboratori Nazionali di Frascati dell'INFN, via. E. Fermi, 40, 
Frascati, Italy} 
\address[d]{Department of Physics, Seoul National University, 151-742 Seoul, 
South Korea}
\address[e]{Dipartimento di Ingegneria Meccanica e Industriale, Universit\`a di 
Brecia, via Valotti 9, Brescia, Italy} 
\address[f]{INFN Sezione di Pavia, via Bassi 6, Pavia, Italy} 
\address[g]{Dipartimento di Fisica, Universit\`a di Torino, 
Via P. Giuria 1, Torino, Italy} 
\address[h]{Dipartimento di Fisica, Universit\`a di Trieste, via Valerio 2, 
Trieste, Italy} 
\address[i]{INFN Sezione di Trieste, via Valerio 2, Trieste, Italy} 
\address[z]{CEA, Irfu/SACM, Gif-sur-Yvette, France} 
%\address[j]{Dipartimento di Fisica Generale, Universit\`a di Torino, 
%Via P. Giuria 1, Torno, Italy} 
\address[k]{Dipartimento di Fisica, Universit\`a di Bari, via Amendola 173, 
Bari, Italy} 
\address[l]{INFN Sezione di Bari, via Amendola 173, Bari, Italy} 
\address[m]{Department of Physics, Kyoto University, Sakyo-ku, Kyoto, Japan} 
\address[n]{Department of Physics, Shahid Behesty University, 19834 Teheran, 
Iran} 
\address[o]{INAF-IFSI, Sezione di Torino, Corso Fiume 4, Torino, Italy} 
\address[p]{RIKEN, Wako, Saitama 351-0198, Japan} 
\address[q]{Racah Institute of Physics, The Hebrew University, 
Jerusalem 91904, Israel} 
\thanks[cor1]{Corresponding author: E. Botta, e-mail address: botta@to.infn.it} 
\thanks[cor2]{deceased} 

\begin{abstract} 
Three candidate events of the neutron-rich hypernucleus $^{6}_{\Lambda}$H 
were uniquely identified in the FINUDA experiment at DA$\Phi$NE, Frascati, 
by observing $\pi^{+}$ mesons from the ($K^{-}_{\rm stop},\pi^{+}$) 
production reaction on $^{6}$Li targets, in coincidence with $\pi^{-}$ 
mesons from $^{6}_{\Lambda}{\rm H}\to {^{6}{\rm He}}+\pi^{-}$ weak decay. 
Details of the experiment and the analysis of its data are reported, 
leading to an estimate of $(2.9\pm 2.0)\cdot 10^{-6}/K^{-}_{\rm stop}$ 
for the $^{6}_{\Lambda}$H production rate times the two-body $\pi^{-}$ 
weak decay branching ratio. The $^{6}_{\Lambda}$H binding energy with 
respect to $^{5}{\rm H}+\Lambda$ was determined jointly from production 
and decay to be $B_{\Lambda}=(4.0\pm 1.1)$ MeV, assuming that $^{5}$H is 
unbound with respect to $^{3}{\rm H}+2n$ by 1.7 MeV. The binding energy 
determined from production is higher, in each one of the three events, 
than that determined from decay, with a difference of $(0.98\pm 0.74)$ 
MeV here assigned to the $0^{+}_{\rm g.s.}\to 1^{+}$ excitation. 
The consequences of this assignment to $\Lambda$ hypernuclear dynamics 
are briefly discussed.  
\end{abstract} 


\end{frontmatter}

\section{Introduction} 

The r$\hat{\mathrm{o}}$le of the $\Lambda$ hyperon in stabilizing nuclear 
cores was pointed out long ago by Dalitz and Levi Setti \cite{dalitz_setti} as 
part of a discussion focusing on light hypernuclei with large neutron excess. 
This property is demonstrated by the observation of $^{6}_{\Lambda}$He, 
$^{7}_{\Lambda}$Be, $^{8}_{\Lambda}$He, $^{9}_{\Lambda}$Be and 
$^{10}_{~\Lambda}$B hypernuclei in emulsion experiments \cite{juric}. 
No unstable-core hydrogen $\Lambda$ hypernuclei have been established so far, 
although the existence of the lightest possible one $^{6}_{\Lambda}$H was  
predicted by Dalitz and Levi Setti \cite{dalitz_setti} and subsequently 
reinforced in estimates by Majling \cite{majling}. The neutral-baryon 
excess in $^{6}_{\Lambda}$H, in particular, would be $(N+Y)/Z=5$, with $Y=1$ 
for a $\Lambda$ hyperon, larger than the maximal value in light nuclei, 
$N/Z=3$ for $^{8}$He \cite{tilley04}. Neutron-rich light hypernuclei could 
thus go beyond the neutron drip line for ordinary nuclear systems. 

Two-body reactions in which neutron rich hypernuclei could be produced 
are the following double charge-exchange reactions: 
\begin{equation} 
K^{-} + {^{A}Z}  \rightarrow {^{A}_{\Lambda}(Z-2)} + \pi^{+}, 
\label{prodK} 
\end{equation} 
induced on nuclear targets by stopped $K^{-}$ mesons or in flight, and 
\begin{equation} 
\pi^{-} + {^{A}Z}  \rightarrow {^{A}_{\Lambda}(Z-2)} + K^{+} 
\label{prodpi} 
\end{equation} 
with $\pi^{-}$ mesons in flight ($p_{\pi^{-}} > 0.89$ GeV/c). 

The simplest description of the above reactions is a two-step process on 
two different protons of the same nucleus, converting them into a neutron 
and a $\Lambda$, with the additional condition that the final nuclear system 
is bound. For (\ref{prodK}) it amounts to $K^{-}p\rightarrow\Lambda\pi^{0}$ 
reaction followed by $\pi^{0}p\rightarrow n\pi^{+}$ or $K^{-}p\rightarrow 
{\bar K^0}n$ followed by ${\bar K^0}p\rightarrow\Lambda\pi^{+}$, 
for (\ref{prodpi}) to a $\pi^{-}p\rightarrow n\pi^{0}$ reaction followed by 
$\pi^{0}p \rightarrow K^{+}\Lambda$ or $\pi^{-}p\rightarrow K^{0}\Lambda$ 
followed by $K^{0}p\rightarrow K^{+}n$. Another mechanism is a single-step 
double charge exchange $m_i^{-}p\rightarrow\Sigma^{-}m_f^{+}$ (where $m$ 
stands for meson) feeding the $\Sigma$ component coherently admixed into the 
final $\Lambda$ hypernuclear state. Such admixtures are essentially equivalent 
to invoking a second step of $\Sigma^{-}p\rightarrow\Lambda n$ conversion. 
These two-step processes are expected to occur at a much lower rate (reduction 
factor $\leq 10^{-2}$ \cite{chrien}) than the production of normal $\Lambda$ 
hypernuclei by means of the corresponding single-step two-body reactions 
($K^{-},\pi^{-}$) and ($\pi^{+},K^{+}$). 

The first experimental attempt to produce neutron-rich hypernuclei by 
the reaction (\ref{prodK}) with $K^{-}$ at rest was carried out at 
KEK \cite{kubota}. Upper limits were obtained for the production of 
$^{9}_{\Lambda}$He, $^{12}_{~\Lambda}$Be and $^{16}_{~\Lambda}$C hypernuclei 
(on $^{9}$Be, $^{12}$C and $^{16}$O targets respectively) in the range of 
$(0.6-2.0)\cdot 10^{-4}/K^{-}_{\rm stop}$, while the theoretical predictions 
for $^{12}_{~\Lambda}$Be and $^{16}_{~\Lambda}$C \cite{tretyak01} lie in the 
interval $(10^{-6}-10^{-7})/K^{-}_{\rm stop}$, which is at least one order 
of magnitude lower than the experimental upper limits and three orders of 
magnitude smaller than the standard one-step ($K^{-}_{\rm stop},\pi^{-}$) 
reaction rates on the same targets ($10^{-3}/K^{-}_{\rm stop}$). 

Another KEK experiment \cite{saha} reported the observation of 
$^{10}_{~\Lambda}$Li in the ($\pi^{-},K^{+}$) reaction on a $^{10}$B target 
with a 1.2 GeV/c $\pi^{-}$ beam. A production cross section of $11.3\pm 1.9$ 
nb/sr was evaluated; the result, however, is not directly comparable with 
theoretical calculations \cite{tretyak03} since no discrete structure was 
observed and the production cross section was integrated over the whole bound 
region ($0<B_{\Lambda}<20$ MeV). 

A further attempt to observe neutron-rich hypernuclei by means of the 
reaction (\ref{prodK}), with $K^{-}$ at rest, was made at the DA$\Phi$NE 
collider at LNF by the FINUDA experiment \cite{nrich1}, on $^{6}$Li and 
$^{7}$Li targets. The limited data sample collected during the first run 
period of the experiment was used to estimate the production rates per 
stopped $K^{-}$ of $^{6}_{\Lambda}$H and $^{7}_{\Lambda}$H. The inclusive 
$\pi^{+}$ spectra from $^{6}$Li and $^{7}$Li targets were analyzed in 
momentum regions corresponding, through momentum and energy conservation, 
to $B_{\Lambda}$ values discussed in the literature. Because of the dominant 
contribution of the reactions 
\begin{eqnarray} 
K^{-}_{\rm stop} + p & \rightarrow & \Sigma^{+} + \pi^{-} \nonumber \\ 
          &             & \hookrightarrow n + \pi^{+} \ \ \ 
(\sim 130 < p_{\pi^{+}} < 250\ {\rm MeV/c}) 
\label{Kp} 
\end{eqnarray} 
and 
\begin{eqnarray} 
K^{-}_{\rm stop} + p p & \rightarrow & \Sigma^{+} + n \nonumber \\ 
            &             & \hookrightarrow n + \pi^{+} \ \ \ 
(\sim 100 < p_{\pi^{+}} < 320\ {\rm MeV/c}), 
\label{Kpp} 
\end{eqnarray} 
which give the main component of the inclusive $\pi^{+}$ spectra for 
absorption of stopped $K^{-}$ mesons on nuclei, and owing to a limited 
statistics, only upper limits could be evaluated for $\Lambda$ hypernuclear 
production: 
\begin{eqnarray} 
R_{\pi^{+}}({^{6}_{\Lambda}{\rm H}}) & < & (2.5 \pm 
{0.4_{\rm stat}}^{+0.4}_{-0.1{\rm syst}})\cdot 10^{-5}/K^{-}_{\rm stop}, 
\label{Rpi+6} \\
R_{\pi^{+}}({^{7}_{\Lambda}{\rm H}}) & < &(4.5 \pm 
{0.9_{\rm stat}}^{+0.4}_{-0.1{\rm syst}})\cdot 10^{-5}/K^{-}_{\rm stop}, 
\label{Rpi+7} \\  
\nonumber 
\end{eqnarray} 
in addition to an upper limit determined in $^{12}$C: 
\begin{eqnarray} 
R_{\pi^{+}}({^{12}_{~\Lambda}{\rm Be}}) & < & (2.0 \pm 
{0.4_{\rm stat}}^{+0.3}_{-0.1{\rm syst}})\cdot 10^{-5}/K^{-}_{\rm stop}, 
\label{Rpi+12} 
\end{eqnarray}  
lowering by a factor $\sim 3$ the previous KEK determination \cite{kubota}. 

In this article we present the analysis of the total data sample 
of the FINUDA experiment, collected from 2003 to 2007 and corresponding to 
a total integrated luminosity of 1156 pb$^{-1}$, aiming at assessing the 
existence of $^{6}_{\Lambda}$H and determining the production rate by means 
of the ($K^{-}_{\rm stop},\pi^{+}$) reaction on $^{6}$Li targets. 
A preliminary account of the results, reporting three clear events of 
$^{6}_{\Lambda}$H, appeared in \cite{PRL}. 

\begin{figure}[htbp] 
\begin{center} 
\includegraphics[width=90mm]{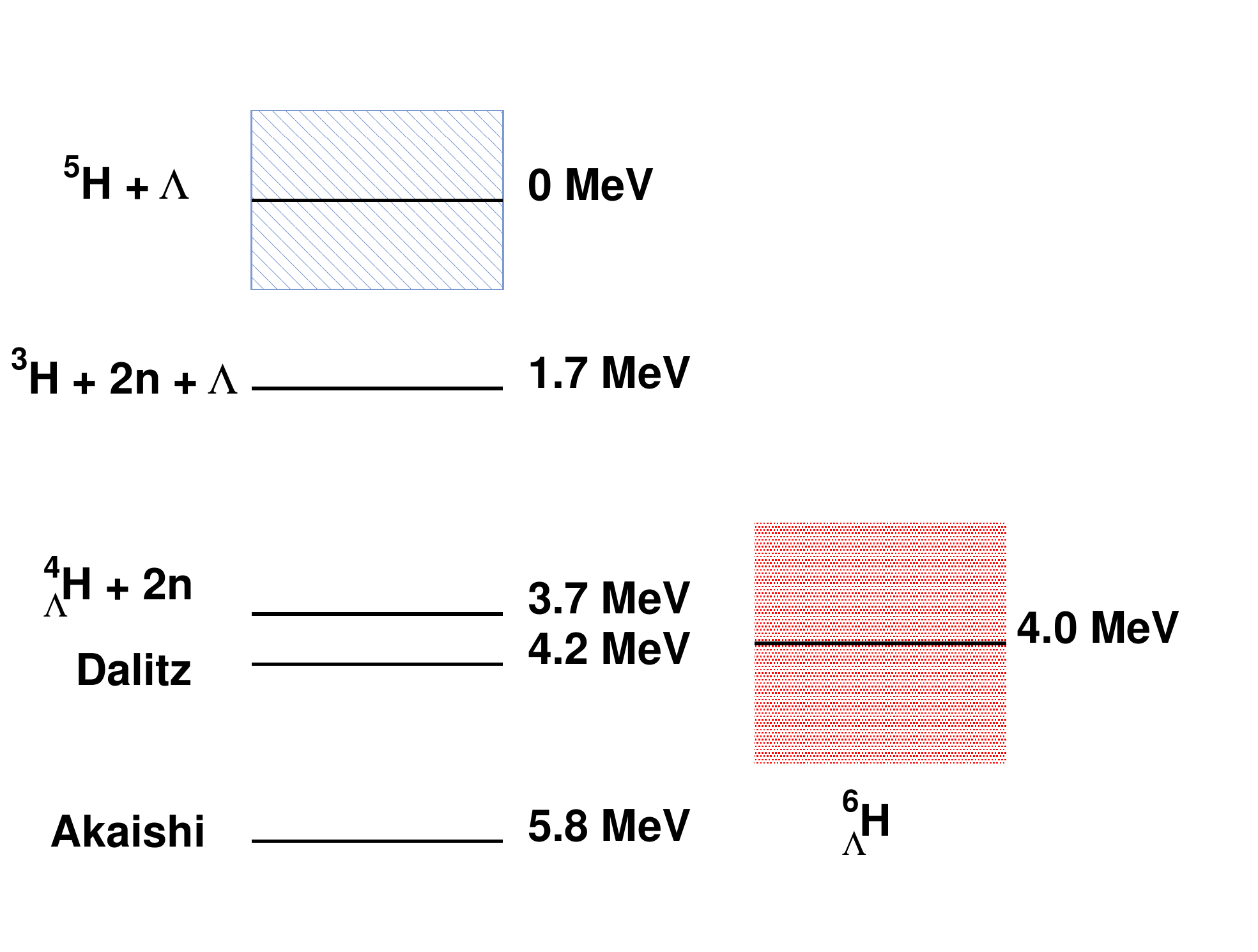} 
\caption{Left: Binding energy scheme for a system of one proton, four neutrons 
and one $\Lambda$ relative to the summed mass of $^{5}{\rm H}+\Lambda$, 
$M=5805.44$ MeV, with the (blue in the web version) hatched box denoting the width of $^{5}$H 
\cite{korshen}. The two lowest horizontal lines stand for predictions from 
Refs.~\cite{dalitz_setti,akaishi}. Right: mean value of the 
$^{6}_{\Lambda}{\rm H}$ g.s. mass obtained in the present analysis jointly 
from production and decay; the (red in the web version) shaded box represents the error on the 
mass mean value obtained from the three $^{6}_{\Lambda}{\rm H}$ events 
reported here.} 
\label{fig6} 
\end{center} 
\end{figure} 

The binding energy of $^{6}_{\Lambda}$H with respect to the unstable $^{5}$H 
core was estimated in Refs.~\cite{dalitz_setti,majling} as $B_{\Lambda}=4.2$ 
MeV, making $^{6}_{\Lambda}$H particle stable with respect to its 
$^{4}_{\Lambda}{\rm H}+2n$ lowest threshold, as shown in Fig.~\ref{fig6}. 
We recall that the binding energy $B_{\Lambda}$ of hypernucleus 
$^{A}_{\Lambda}$Z is defined as: 
\begin{equation} 
B_{\Lambda} = M_{\rm core} + M_{\Lambda} - M_{^{A}_{\Lambda}{\rm Z}}, 
\label{blambda} 
\end{equation} 
where $M_{\rm core}$ is the mass of the $^{(A-1)}$Z core nucleus in its 
ground state (g.s.), as deduced from the atomic mass tables \cite{wapstra}. 
The $^{5}$H nuclear core, colloquially termed ``superheavy hydrogen", was 
observed as a broad resonance (1.9 MeV FWHM) at energy about 1.7 MeV above 
the $^{3}{\rm H}+2n$ threshold \cite{korshen}. A substantially stronger 
binding, $B_{\Lambda}({^{6}_{\Lambda}{\rm H}})=5.8$ MeV, was predicted by 
Akaishi et al. \cite{akaishi} for the $0^+$ g.s. on the basis of a coherent 
$\Lambda N - \Sigma N$ mixing model originally practised for the 
$^{4}_{\Lambda}$H cluster \cite{akaishi00}. This coherent $\Lambda N-\Sigma N$ 
mixing induces a spin-dependent $\Lambda NN$ three-body interaction which 
affects primarily the $0^+$ g.s., increasing thus the $\approx$ 1 MeV $1^+$ 
excitation expected from $^{4}_{\Lambda}$H to 2.4 MeV in $^{6}_{\Lambda}$H. 
If this prediction is respected by Nature, it could imply far-reaching 
consequences to strange dense stellar matter. 

In the next sections we describe briefly the FINUDA experimental apparatus, 
and the analysis technique applied to the data collected on $^{6}$Li targets. 
We then report on three $^{6}_{\Lambda}$H candidate events found by observing 
$\pi^+$ mesons from production and $\pi^-$ mesons from decay in coincidence. 
These events prove robust against varying the cuts selected in the analysis, 
and give evidence for a particle stable $^{6}_{\Lambda}$H. The measurement 
background is evaluated and the production rate of $^{6}_{\Lambda}$H is 
estimated. We end with a brief discussion of the $^{6}_{\Lambda}$H excitation 
spectrum as constrained by the three candidate events.

\section{Experimental apparatus} 

FINUDA was a hypernuclear physics experiment installed at one of the two 
interaction regions of the DA$\Phi$NE $e^{+}e^{-}$ collider, the INFN-LNF 
$\Phi$(1020)-factory. A detailed description of the experimental apparatus 
can be found in Ref.~\cite{fnd}. The layout figured a cylindrical symmetry 
arrangement; here we briefly sketch its main components moving outwards 
from the beam axis: the {\it interaction/target region}, composed by a barrel 
of 12 thin scintillator slabs (TOFINO), surrounded by an octagonal array of 
$\mathrm{Si}$ microstrips (ISIM) facing eight target tiles; the {\it tracking 
device}, consisting of four layers of position sensitive detectors 
(a decagonal array of $\mathrm{Si}$ microstrips (OSIM), two octagonal layers 
of low mass drift chambers (LMDC) and a stereo system of straw tubes (ST)) 
arranged in coaxial geometry; the {\it external time of flight detector} 
(TOFONE), a barrel of 72 scintillator slabs. The whole apparatus was placed 
inside a uniform 1.0 T solenoidal magnetic field; the tracking volume was 
immersed in $\mathrm{He}$ atmosphere to minimize the multiple scattering 
effect. 

The main features of the apparatus were the thinness of the target 
materials needed to stop the low energy ($\sim$ 16 MeV) $K^{-}$'s from 
the $\Phi\rightarrow K^{-}K^{+}$ decay channel, the high transparency of 
the FINUDA tracker and the very large solid angle ($\sim 2\pi$ sr) covered 
by the detector ensemble; accordingly, the FINUDA apparatus was suitable 
to study simultaneously the formation and the decay of $\Lambda$ hypernuclei 
by means of high resolution magnetic spectroscopy of the emitted charged 
particles. 

In particular, for $\pi^{+}$ with momentum $\sim 250$ MeV/c the resolution 
of the tracker can be evaluated by measuring the width of the momentum 
distribution of the monochromatic (235.6 MeV/c) $\mu^{+}$ coming from the 
$K_{\mu 2}$ decay channel; for reactions occurring in the apparatus sector 
where $^{6}$Li targets were located, it is $\sigma _{p} = (1.1 \pm 0.1)$ 
MeV/c \cite{spectrFND}; the precision on the absolute momentum calibration, 
obtained from the mean value of the same distribution, is better than 0.12 
MeV/c for the $^{6}$Li targets, which corresponds to a maximum systematic 
uncertainty in the kinetic energy $\sigma_{T sys}(\pi^{+}) = 0.1$ MeV. 

\begin{figure}[htbp] 
\begin{center} 
\includegraphics[width=90mm]{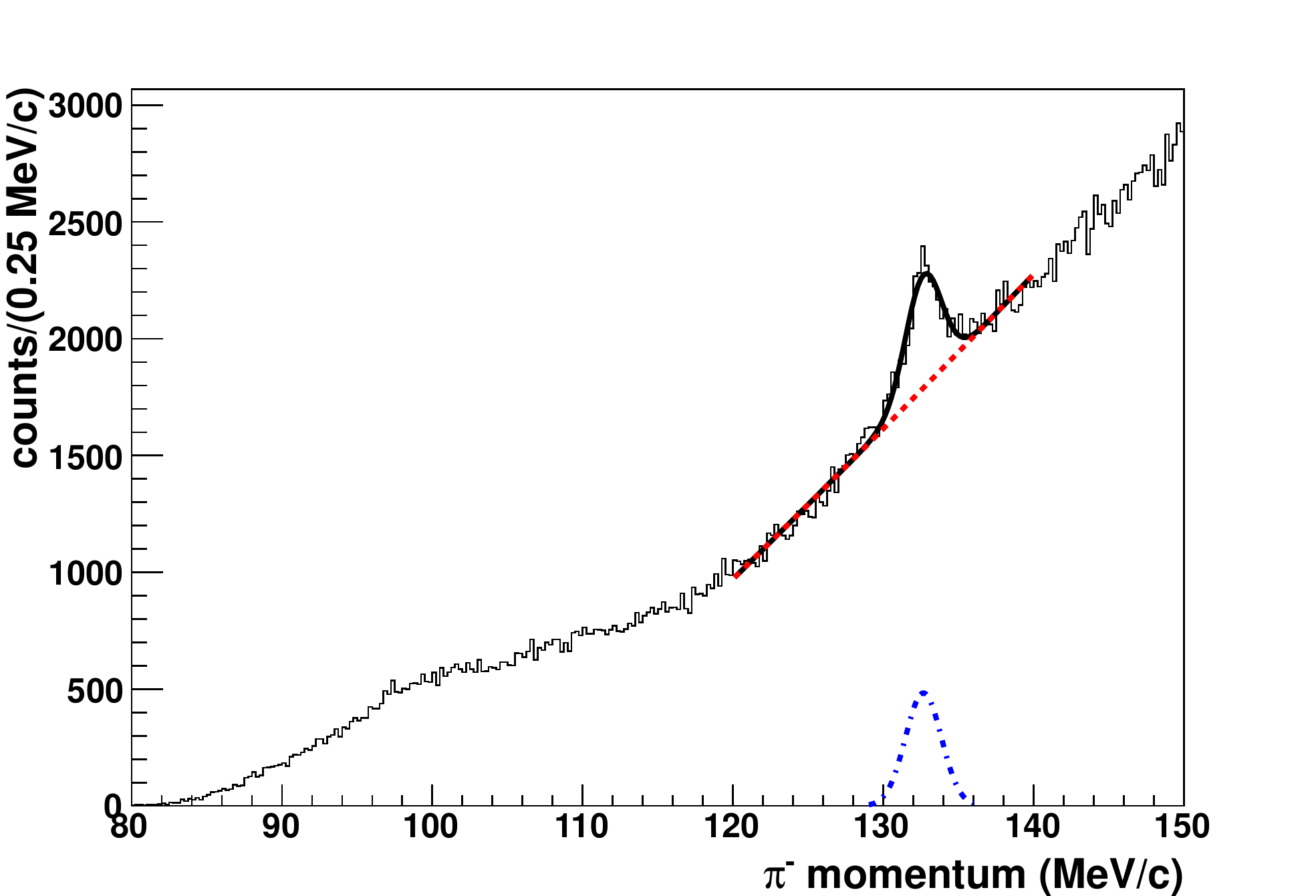} 
\caption{Distribution of low momentum $\pi^{-}$ from $^{6}$Li targets. 
The continuous black curve represents a fit the the spectrum given by 
the sum of a second degree polynomial [dashed (red in the web version) curve] and a gaussian 
function [dot-dashed (blue in the web version) curve]. For more details, see text.} 
\label{fig1} 
\end{center} 
\end{figure} 

For $\pi^{-}$ with momentum $\sim 130$ MeV/c the resolution and absolute 
calibration can be evaluated from the momentum distribution of the 
monochromatic $\pi^{-}$ coming from the two-body mesonic weak decay of 
$^{4}_{\Lambda}$H, produced as hyperfragment with a formation probability 
of the order of $10^{-3}-10^{-2}$ per stopped $K^{-}$ \cite{tamura_fr}. 
Figure~\ref{fig1} shows the distribution for low momentum $\pi^{-}$ from 
$^{6}$Li targets, before acceptance correction; the spectrum is fitted in 
the 120--140 MeV/c momentum range (continuous black curve) with the sum of 
a second degree polynomial, representing the background from quasi-free 
$\Lambda$ decay and quasi-free $\Sigma^{+}$ production (dashed (red in the web version) curve 
in the figure), and a gaussian function representing the $^{4}_{\Lambda}$H 
mesonic decay contribution (dot-dashed (blue in the web version) curve); the fit gives 
a $\chi^{2}/{\rm ndf}=79.1/74$, a mean $\mu_{p}=(132.6\pm 0.1)$ MeV/c and 
a standard deviation $\sigma _{p}= (1.2 \pm 0.1)$ MeV/c for the gaussian 
function, directly measuring the experimental resolution. 
For comparison, $p_{\pi^-}=(132.80\pm 0.08)$ MeV/c from 
$B_{\Lambda}({_{\Lambda}^{4}{\rm H}})=2.04\pm 0.04$ MeV, 
as determined from emulsion studies \cite{juric};  hence the absolute uncertainty is 0.2 MeV/c. 
and the corresponding systematic uncertainty in the kinetic energy is 
then $\sigma_{T sys}(\pi^{-})=0.14$ MeV. 

To perform particle identification, the information of the specific energy 
loss in both OSIM and the LMDC's is used; the mass identification from the 
time of flight system (TOFINO-TOFONE) for high momentum tracks is also used. 
The final selection is performed by requiring the same identification from 
at least two different detectors. 

\section{Analysis technique} 

In the second data taking the statistics collected with $^{6}$Li targets was 
improved by a factor 5 with respect to the first run. However, even with the 
improved statistics, we could not observe in the inclusive $\pi^{+}$ spectra 
clear peaks that could be attributed to the two-body reaction: 
\begin{equation} 
K^{-}_{\rm stop}+{^{6}{\rm Li}}\rightarrow {^{6}_{\Lambda}{\rm H}}+\pi^{+}\ \ \ 
(p_{\pi^{+}}\sim 252\ \mathrm{MeV/c}). 
\label{nrich6LH} 
\end{equation} 
Exploiting the increased statistics, we tried then to reduce the background 
overwhelming the events from reaction (\ref{nrich6LH}) by examining the 
spectra of $\pi^{+}$ in coincidence with the $\pi^{-}$ coming from the mesonic 
decay of $^{6}_{\Lambda}$H: 
\begin{equation} 
{^{6}_{\Lambda}{\rm H}} \rightarrow {^{6}{\rm He}} + \pi^{-} \ \ \ 
(p_{\pi^{-}}\sim 130-140\ \mathrm{MeV/c}). 
\label{6LHmwd} 
\end{equation} 
The branching ratio for (\ref{6LHmwd}) is expected to be about 50$\%$ 
taking into account the value measured for the analogous decay 
${^{4}_{\Lambda}{\rm H}}\rightarrow {^{4}{\rm He}}+\pi^{-}$ \cite{tamura_fr}. 
($\pi^{+}$, $\pi^{-}$) coincidence events, associated with $K^{-}$'s 
stopped in the $^{6}$Li targets, were thus considered; only reaction 
(\ref{Kp}) contributes to the background of this sample. 

\begin{figure}[htbp] 
\begin{center} 
\includegraphics[width=90mm]{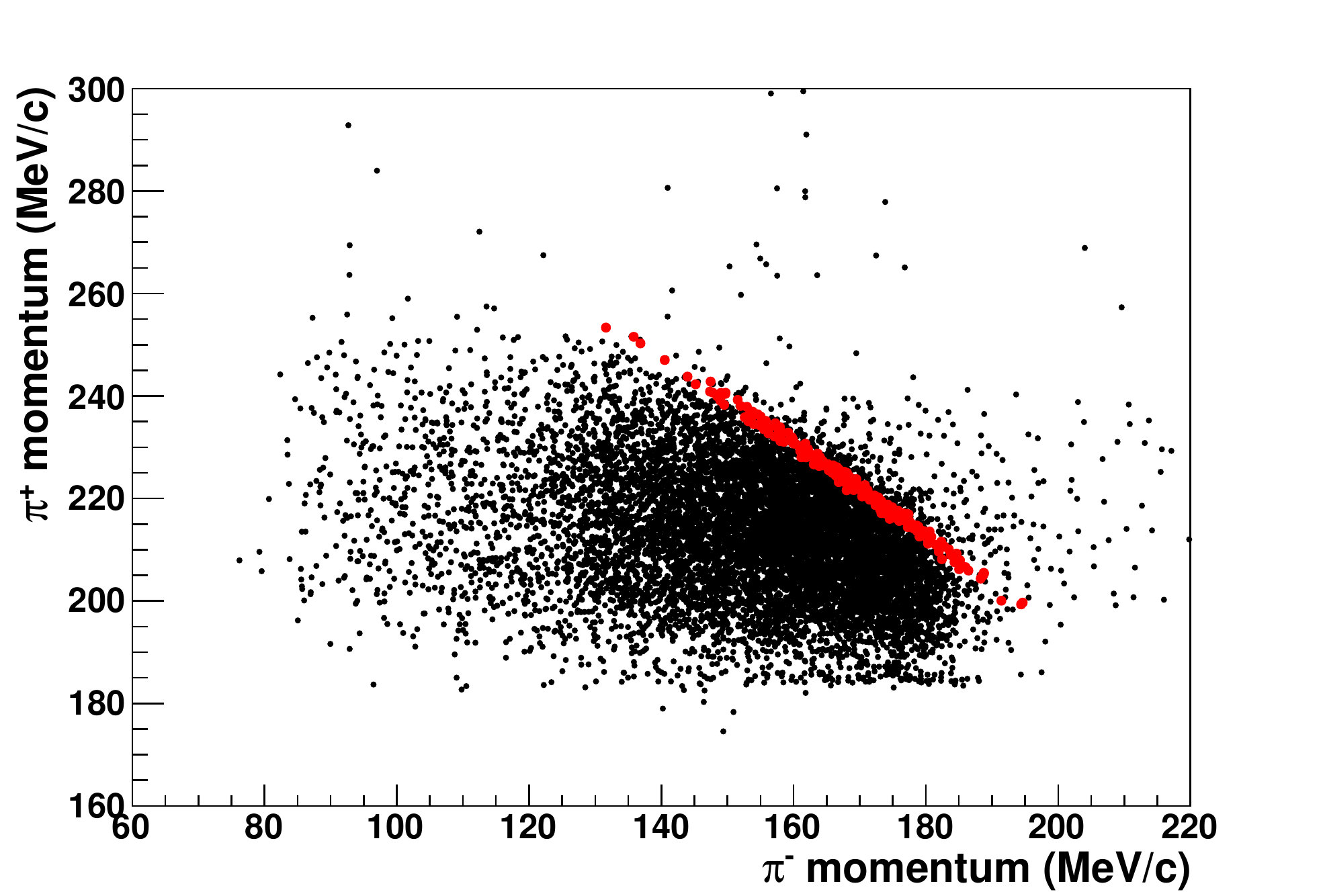} 
\caption{$\pi^{+}$ momentum vs $\pi^{-}$ momentum for $^{6}$Li targets. 
The (red in the web version) band dots stand for events with $T(\pi^{+})+T(\pi^{-})=202-204$ MeV. 
See text for more details.} 
\label{fig2} 
\end{center} 
\end{figure}

We examined thus the two-dimensional raw spectrum of $\pi^{+}$ versus 
$\pi^{-}$ momentum, shown in Fig.~\ref{fig2}, in order to recognize 
possible enhancements due to occurrence of  the reactions (\ref{nrich6LH}) 
and (\ref{6LHmwd}) in sequence. The low statistics and the strong background 
prevented us from finding statistically significant accumulations of events 
in the plot arising from a bound $^{6}_{\Lambda}$H. 

In order to isolate the events due to the possible formation of a bound 
$^{6}_{\Lambda}$H, we considered energy conservation for both reactions 
(\ref{nrich6LH}) and (\ref{6LHmwd}). Momentum conservation is automatically 
ensured by the fact that both reactions of formation (\ref{nrich6LH}) and 
decay (\ref{6LHmwd}) occur at rest. The stopping time of $^{6}_{\Lambda}$H 
in the material is indeed shorter than its lifetime. 

For (\ref{nrich6LH}) we may write explicitly: 
\begin{equation} 
M(K^{-})+3M(p)+3M(n)-B(^{6}{\rm Li})=M({^{6}_{\Lambda}{\rm H}})+
T({^{6}_{\Lambda}{\rm H}})+M(\pi^{+})+T(\pi^{+}), 
\label{Eform} 
\end{equation} 
in which, in obvious notation, $M$ stands for a particle mass, $T$ -- its 
kinetic energy, and $B({^{6}{\rm Li}})$ -- the binding energy of $^{6}$Li. 
For (\ref{6LHmwd}) we may write: 
\begin{equation}
M({^{6}_{\Lambda}{\rm H}})=2M(p)+4M(n)-B({^{6}{\rm He}})+T({^{6}{\rm He}})+
M(\pi^{-})+T(\pi^{-}), 
\label{Edecay} 
\end{equation} 
in the same notation as above. Combining Eqs.~(\ref{Eform}) and (\ref{Edecay}) 
in order to eliminate $M({^{6}_{\Lambda}{\rm H}})$, we get the following 
equation: 
\begin{eqnarray} 
T(\pi^{+}) + T(\pi^{-}) & = &  M(K^{-}) + M(p) - M(n) - 2M(\pi) \nonumber \\ 
& & - B({^{6}{\rm Li}}) + B({^{6}{\rm He}}) -T({^{6}{\rm He}})-
T({^{6}_{\Lambda}{\rm H}}). 
\label{Ebal} 
\end{eqnarray} 
All the terms on the right-hand side are either known constants or quantities 
that can be evaluated from momentum and energy conservation, except for 
$T({^{6}_{\Lambda}{\rm H}})$ ($T({^{6}{\rm He}})$) that depends explicitly 
(implicitly) on the unknown value of $B_{\Lambda}({^{6}_{\Lambda}{\rm H}})$. 
A variation of $B_{\Lambda}({^{6}_{\Lambda}{\rm H}})$ between 0 and 6 MeV 
introduces a change of $\sim$ 0.3 MeV in the kinetic energy 
$T({^{6}_{\Lambda}{\rm H}})$ in (\ref{Eform}), corresponding to a sensitivity 
of 50 keV per MeV of $B_{\Lambda}({^{6}_{\Lambda}{\rm H}})$, and a change of 
$\sim 0.2$ MeV in $T(\pi^{+})+T(\pi^{-})$ in (\ref{Ebal}), corresponding to 
a sensitivity of 30 keV per MeV of $B_{\Lambda}({^{6}_{\Lambda}{\rm H}})$. 
These variations are much lower than the experimental energy resolutions for 
$\pi^{+}(250\ \mathrm{MeV/c})$ and $\pi^{-}(130\ \mathrm{MeV/c})$: 
$\sigma_{T}(\pi^{+})=0.96$ MeV and $\sigma_{T}(\pi^{-})=0.84$ MeV. The FINUDA 
energy resolution for a ($\pi^{+},\pi^{-}$) pair in coincidence is therefore 
$\sigma_{T}=\sqrt{\sigma_{T{\rm exp}}^{2}+\sigma_{T{\rm sys}}^{2}}=1.3$ MeV, 
where $\sigma_{T{\rm exp}}=\sqrt{0.96^{2}+0.84^{2}}=1.3$ MeV is 
the total experimental energy resolution and $\sigma_{T{\rm sys}}=
\sqrt{\sigma_{T{\rm sys}}(\pi^{+})^{2}+\sigma_{T{\rm sys}}(\pi^{-})^{2}}=0.17$ 
MeV is the total systematic error on energy. To be definite, we assume 
a value of $B_{\Lambda}({^{6}_{\Lambda}{\rm H}})=5$ MeV, halfway between 
the conservative estimate of 4.2 MeV \cite{dalitz_setti,majling} and Akaishi's 
prediction of 5.8 MeV \cite{akaishi}. The r.h.s. of Eq.~(\ref{Ebal}) assumes 
then a value of $T(\pi^{+})+T(\pi^{-})=203.0\pm 1.3$ MeV. 

\begin{figure}[htbp] 
\begin{center} 
\includegraphics[width=80mm]{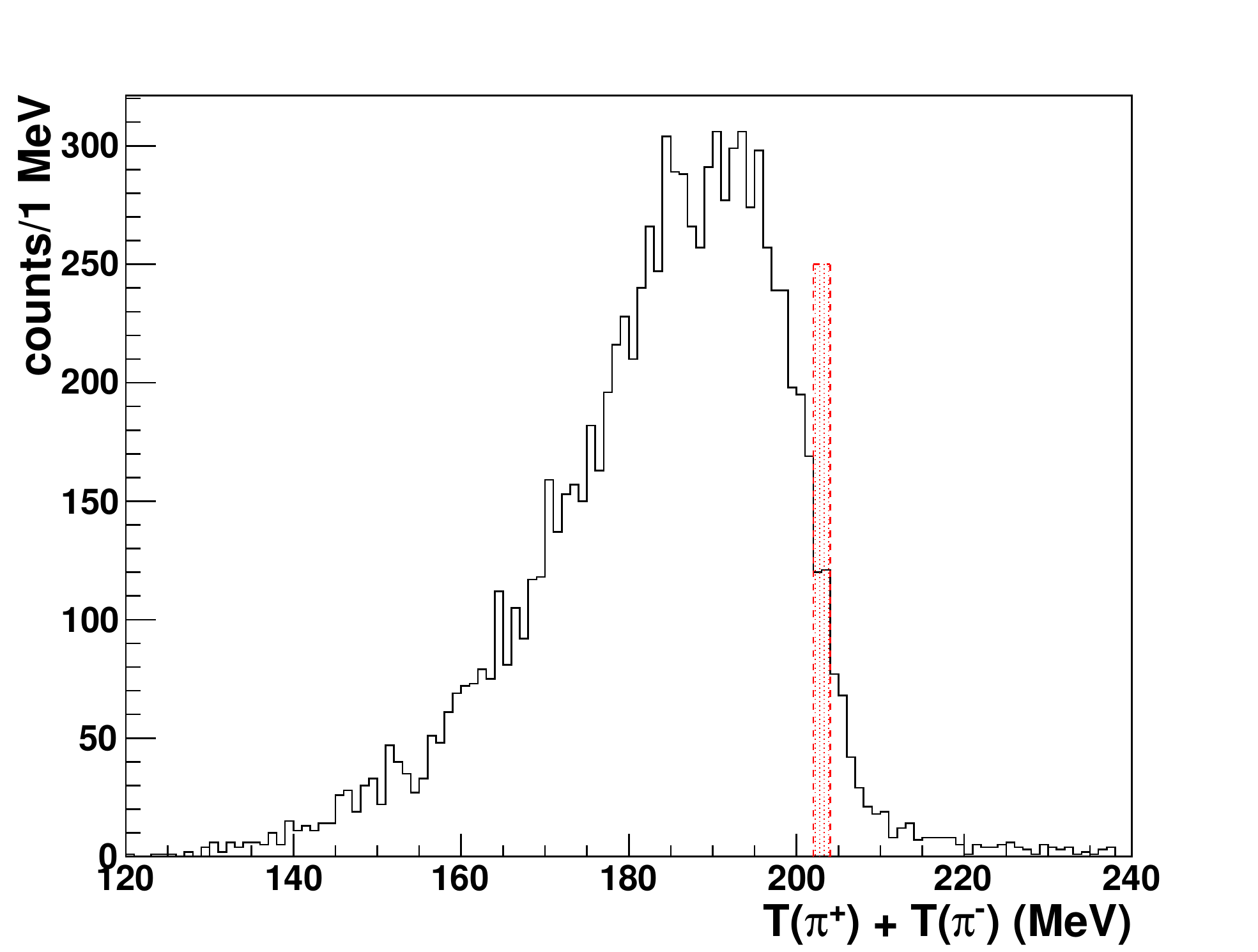} 
\caption{Distribution of raw total kinetic energy $T(\pi^{+})+T(\pi^{-})$ 
for ($\pi^{+},\pi^{-}$) coincidence events from $^{6}$Li targets. The (red in the web version) 
shaded vertical bar represents the cut $T(\pi^{+})+T(\pi^{-})=202-204$ MeV.} 
\label{fig3} 
\end{center} 
\end{figure} 
\begin{figure}[htbp] 
\begin{center} 
\includegraphics[width=80mm]{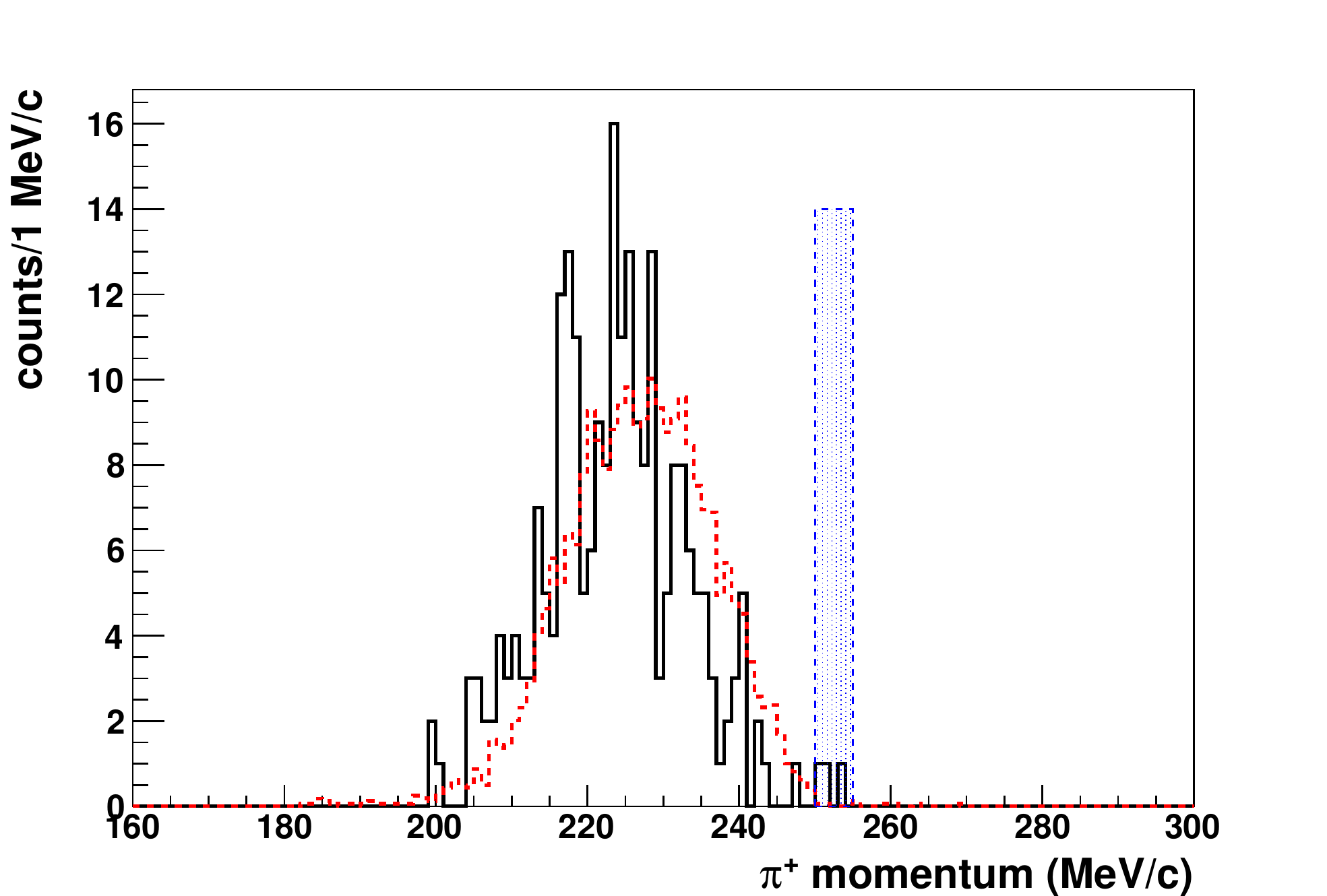} 
\hspace{2mm} 
\includegraphics[width=80mm]{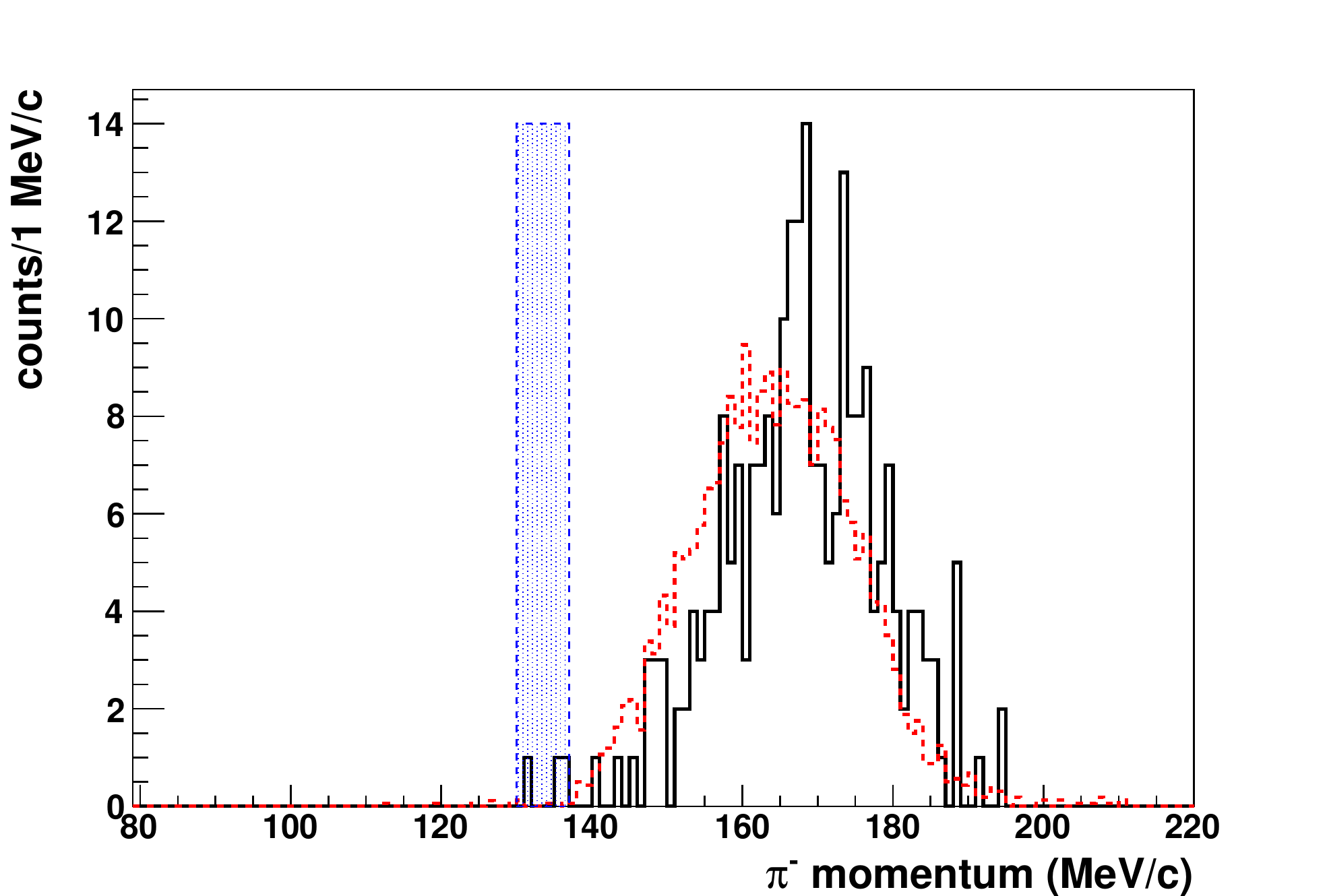} 
\caption{Continuous histograms: distribution of $\pi^{+}$ 
(upper part) and $\pi^{-}$ (lower part) momenta for ($\pi^{+},\pi^{-}$) 
coincidence events with $T(\pi^{+})+T(\pi^{-})=202-204$ MeV from $^{6}$Li 
targets before acceptance correction. The (blue in the web version) shaded vertical bars 
indicate the final selection regions. Dashed (red in the web version) histograms represent the 
$\Sigma^{+}$ background spectra: see text for more details.} 
\label{fig4} 
\end{center} 
\end{figure} 

We considered then the raw spectrum of the total kinetic energy, 
$T(\pi^{+})+T(\pi^{-})$, for the coincidence events, shown in Fig.~\ref{fig3}. 
Events in the summed energy distribution were selected in the region 
($203\pm 1$) MeV, indicated by the (red in the web version) filling in the figure. 
The half-width of the interval corresponds to $\sim 77\%$ of the FINUDA 
total energy resolution; the value was chosen as a compromise between the 
strong requirement of reducing the contamination from background reactions, 
as will be discussed in more detail in the following, and the plight for 
reasonable statistics, leading to application of a selection narrower than 
the experimental resolution. The selected events are represented by red dots 
in Fig.~\ref{fig2}. 

The raw distributions of $p_{\pi^{+}}$ and $p_{\pi^{-}}$ for the events 
selected are shown in Fig.~\ref{fig4} by the continuous line histogram, 
falling off to zero at $p_{\pi^{+}}=245$ MeV/c in the higher momentum region, 
and at $p_{\pi^{-}}=145$ MeV/c in the lower momentum region. These limiting 
values, when inserted in Eqs.~(\ref{nrich6LH}) and (\ref{6LHmwd}) for 
two-body kinematics $^{6}_{\Lambda}$H production from rest and decay at rest, 
yield $^{6}_{\Lambda}$H mass values higher than the total mass of both 
($\Lambda+{^{3}\mathrm{H}}+2n$) and ($\Lambda+{^{5}\mathrm{H}}$) thresholds 
marked in Fig.~\ref{fig6}. A $^{6}_{\Lambda}$H mass equal to the mass of its 
lowest particle stability threshold ${^{4}_{\Lambda}{\rm H}}+2n$ corresponds 
to values of $p_{\pi^{+}}=251.9$ MeV/c and $p_{\pi^{-}}=135.6$ MeV/c. 
A genuinely bound $^{6}_{\Lambda}$H system, therefore, requires that pion 
momenta satisfying $p_{\pi^{+}}>251.9$ MeV/c and $p_{\pi^{-}}<135.6$ MeV/c 
are selected. The cuts actually applied in the analysis of the data, 
$p_{\pi^{+}}=(250-255)$ MeV/c and $p_{\pi^{-}}=(130-137)$ MeV/c, as marked by 
the (blue in the web version) shaded vertical bars in Fig.~\ref{fig4}, allow for a wide range of 
$^{6}_{\Lambda}$H masses from the ($\Lambda+{^{3}\mathrm{H}}+2n$) threshold, 
about 2 MeV in the $^{6}_{\Lambda}$H continuum, down to 
$B_{\Lambda}({^{6}_{\Lambda}{\rm H}})\geq 6$ MeV, somewhat below the 
mass predicted by Akaishi \cite{akaishi}. These cuts do not exclude 
completely an eventual contribution from the production and decay of 
${^{4}_{\Lambda}\mathrm{H}} + 2n$, of a weight which is anyway negligible, 
as discussed in the next section.

\section{Results} 

\begin{figure}[htbp] 
\vspace{-5mm} 
\begin{center} 
\includegraphics[width=140mm]{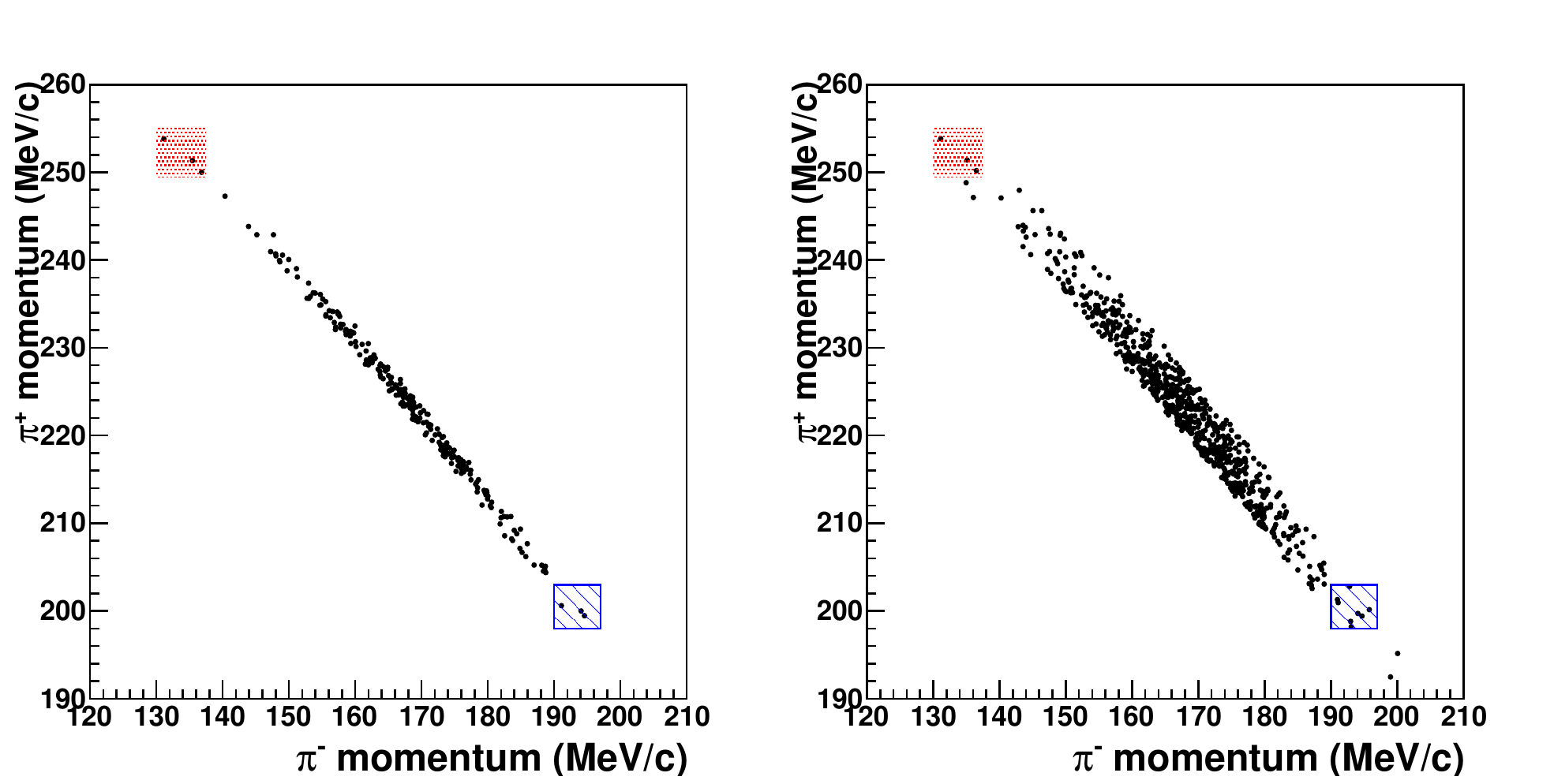} 
\caption{$\pi^{+}$ momentum vs $\pi^{-}$ momentum for $^{6}$Li target 
events with $T(\pi^{+})+T(\pi^{-})=202-204$ MeV (l.h.s.) and with 
$T(\pi^{+})+T(\pi^{-})=200-206$ MeV (r.h.s.). The shaded (red in the web version) rectangles 
on each side consist of a subset of events with $p_{\pi^{+}}=250-255$ MeV/c 
and $p_{\pi^{-}}=130-137$ MeV/c. The hatched (blue in the web version) rectangles on each side 
are symmetric subsets of events to those in the shaded rectangles.} 
\label{fig5} 
\end{center} 
\end{figure}

Three events, out of a total number of $\sim 2.7\cdot 10^7$ $K^{-}$ 
detected at stop on the $^{6}$Li targets, satisfy the final requirements, 
$T(\pi^{+})+T(\pi^{-})=202-204$ MeV, $p_{\pi^{+}}=250-255$ MeV/c and 
$p_{\pi^{-}}=130-137$ MeV/c. These events, within the (red in the web version) shaded rectangle 
on the l.h.s. of Fig.~\ref{fig5}, are candidates for $^{6}_{\Lambda}$H. 
The $\pi^{+}$ momenta which this rectangle encompasses go up from a value 
corresponding to the ($\Lambda+{^{3}\mathrm{H}}+2n$) threshold to a value 
corresponding to the binding energy predicted by Akaishi, whereas the 
$\pi^{-}$ momenta which the rectangle encompasses go down from a value 
corresponding to the same ($\Lambda+{^{3}\mathrm{H}}+2n$) threshold to about 
$2\sigma(p_{\pi^{-}})$ below the value predicted by Akaishi \cite{akaishi}. 

Different choices of $T(\pi^{+})+T(\pi^{-})$ interval widths ($2-6$ MeV) 
and position (center in $202-204$ MeV) and of $p_{\pi^{+}}/p_{\pi^{-}}$ 
interval widths ($5-10$ and $8-15$ MeV/c) with fixed limits at 250 and 
137 MeV/c respectively to exclude the unbound region, affect the populations 
of the corresponding single spectra but not the coincidence spectrum. 
As an example, in Fig.~\ref{fig5} a comparison is made between the 
($p_{\pi^{+}},p_{\pi^{-}}$) plots satisfying the actual selection 
$T(\pi^{+})+T(\pi^{-})=202-204$ MeV (l.h.s.), and similar plots admitting 
a wider selection range $T(\pi^{+})+T(\pi^{-})=200-206$ MeV (r.h.s.). The 
global population increases for the wider cut, as expected, but the events 
that satisfy simultaneously also the separate selections imposed on 
$p_{\pi^{+}}$ and $p_{\pi^{-}}$ (shaded rectangles in the upper left part of 
the plots) remain the same. A similar stability is {\it not} observed in the 
opposite corner of the plots where, on top of the events already there on the 
left plot, five additional events appear on the right plot upon extending the 
cut. Quantitatively, fitting the projected $\pi^{\pm}$ distributions of the 
l.h.s. of Fig.~\ref{fig5} by gaussians, an excess of three events in both 
$p_{\pi^{\pm}}$ distributions is invariably found, corresponding to the shaded 
(red in the web version) rectangle. The probability for the three events to belong to the fitted 
gaussian (background) distribution is less than $0.5\%$ in both cases. 
It is possible, moreover, to see directly from the two dimesional plots that 
variations of the independent momentum selections do not produce any effect. 
Systematic errors due to the applied analysis selection are thus ruled out. 

\begin{figure}[h] 
\begin{center} 
\includegraphics[width=70mm]{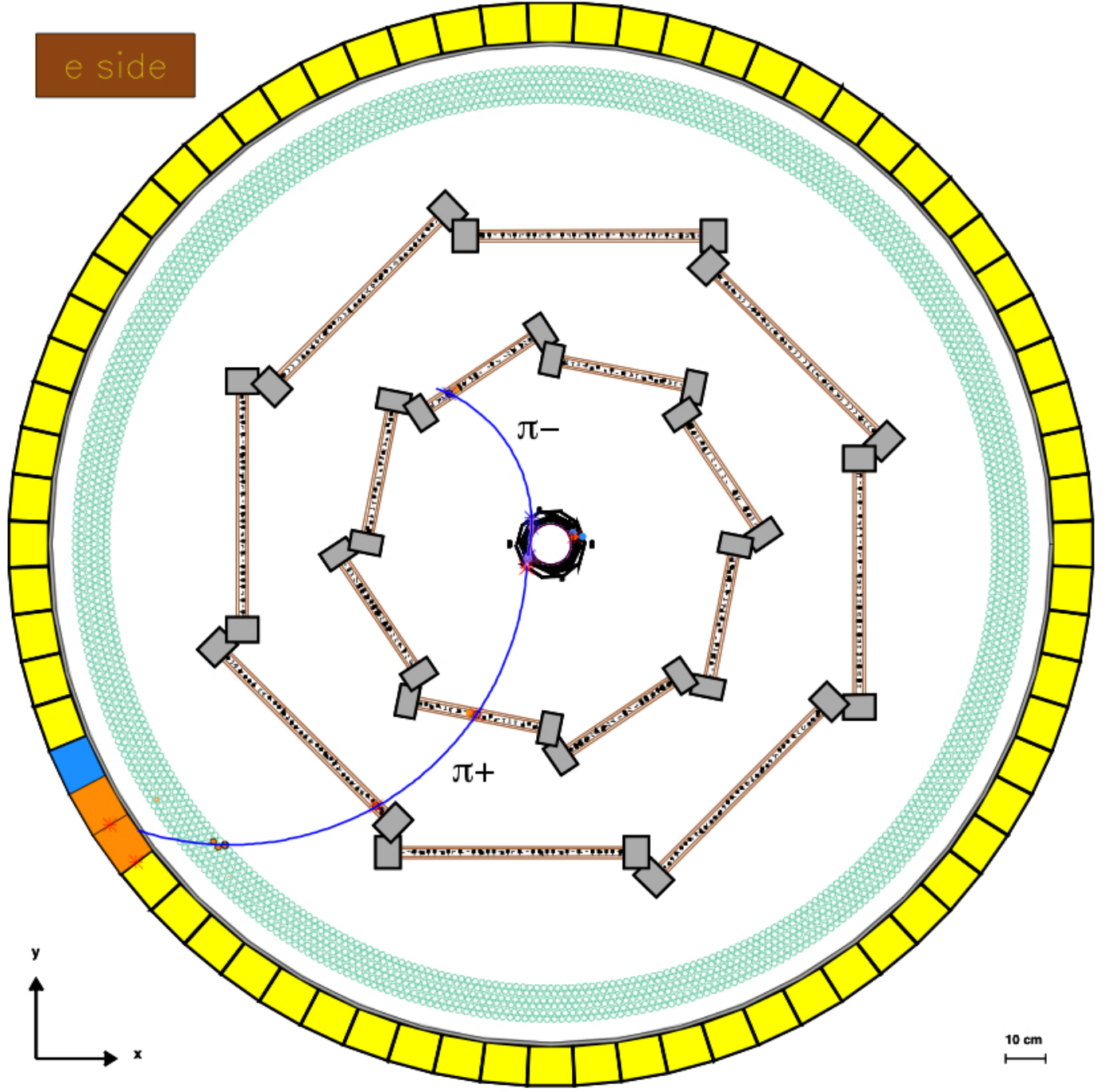} 
\hspace{2mm} 
\includegraphics[width=40mm]{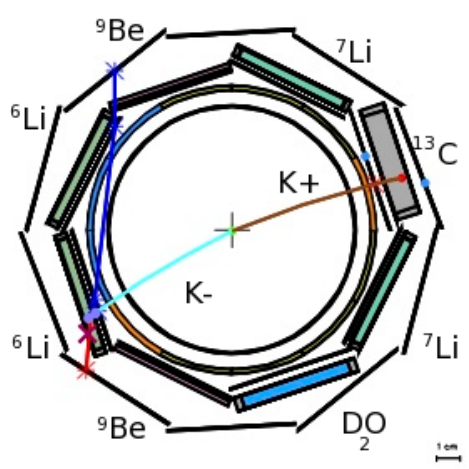} 
\caption{Left: front view of one of the $^{6}_{\Lambda}$H candidate events 
reconstructed by FINUDA where a ($\pi^{+},\pi^{-}$) pair emerges from a 
$^{6}$Li target and crosses the spectrometer. Right: expanded view of the 
target region for the same event where the $K^{-}$ track stops in a $^{6}$Li 
target.} 
\label{fig6b} 
\end{center} 
\end{figure}

It is also worth noticing that the tight momentum cuts imposed on the 
($\pi^{+},\pi^{-}$) coincidence events allow to eliminate completely any 
contamination due to possible $\pi^{-}/e^{-}$ misidentification. Furthermore, 
$\mu^{+}$'s from $K^{+}_{\mu 2}$ decay are clearly separated from $\pi^{+}$'s 
coming from the opposite $K^{-}$ interaction vertex. Figure~\ref{fig6b} shows 
a front view of one of the three events, as reconstructed by FINUDA. 

\begin{table}[h] 
\begin{center} 
\caption{Kinematical properties and $^{6}_{\Lambda}$H mass, 
$M({^{6}_{\Lambda}{\rm H}})$, of the three $^{6}_{\Lambda}$H candidate 
events from production (\ref{nrich6LH}) and decay (\ref{6LHmwd}) reactions. 
Listed in the last two columns are the mean and the difference of the 
production and decay masses. $T_{\rm tot}$ indicates the total $\pi^{\pm}$ 
kinetic energy $T(\pi^{+})+T(\pi^{-})$.}
\label{tab1}
\vspace{2mm} 
\begin{tabular}{|c|c|c|c|c|c|c|} 
%\begin{tabular}{ccccccc} 
\hline 
$T_{\rm tot}$ & $p_{\pi^{+}}$ & $p_{\pi^{-}}$ & $M({^{6}_{\Lambda}{\rm H}})$ & 
$M({^{6}_{\Lambda}{\rm H}})$ & $M({^{6}_{\Lambda}{\rm H}})$ & 
$\Delta M({^{6}_{\Lambda}{\rm H}})$ \\ 
(MeV) & (MeV/c) & (MeV/c) & prod. (MeV) & decay (MeV) & mean (MeV) & (MeV) \\ 
\hline 
202.6$\pm$1.3 & 251.3$\pm$1.1 & 135.1$\pm$1.2 & 5802.33$\pm$0.96 & 
5801.41$\pm$0.84 & 5801.87$\pm$0.96 & 0.92$\pm$1.28 \\ 
%\hline 
202.7$\pm$1.3 & 250.1$\pm$1.1 & 136.9$\pm$1.2 & 5803.45$\pm$0.96 & 
5802.73$\pm$0.84 & 5803.09$\pm$0.96 & 0.72$\pm$1.28 \\ 
%\hline 
202.1$\pm$1.3 & 253.8$\pm$1.1 & 131.2$\pm$1.2 & 5799.97$\pm$0.96 & 
5798.66$\pm$0.84 & 5799.32$\pm$0.96 & 1.31$\pm$1.28 \\ 
\hline 
\end{tabular} 
\end{center} 
\end{table} 

By evaluating event by event the corresponding $^{6}_{\Lambda}$H mass from 
both production (\ref{nrich6LH}) and decay (\ref{6LHmwd}) reactions, the mass 
values listed in Table~\ref{tab1} are obtained. A mean value for each event 
jointly from production and decay was also evaluated, with an error given by 
the highest of the two uncertainties, 0.96 MeV. For the global mean mass value 
of the three events we then find 
${\bar M}({^{6}_{\Lambda}{\rm H}})=(5801.43\pm 0.55)$ MeV, 
where the uncertainty $0.55=0.96/\sqrt{3}$ MeV reflects the uncertainty 
assigned to each event. This global mass uncertainty, however, is considerably 
smaller than the mean-mass spread of the three events. We therefore decided to 
relax the assigned uncertainty by calculating it from the spread of the three 
mean mass values, which yields 
\begin{equation}
{\bar M}({^{6}_{\Lambda}{\rm H}}) = 5801.4\pm 1.1~{\rm MeV}, 
\label{finalmass} 
\end{equation} 
with uncertainty larger than the 0.96 MeV and 0.84 MeV measurement 
uncertainties in production and decay respectively. The standard deviation 
of this uncertainty is 0.55 MeV which together with $\sigma=1.1$ MeV is still 
short of the 2.11 MeV deviation of the third event mean mass from the mean 
mass value. This observation could indicate some irregularity in the 
reconstruction of the third event. To regain confidence, each one of the 
three events was checked visually for irregularities but none was found. 
The third event, in particular, is shown in Fig.~\ref{fig6b}. 

Listed in the last column of Table~\ref{tab1} are values of 
$\Delta M({^{6}_{\Lambda}{\rm H}})$, defined as the difference between the 
$^{6}_{\Lambda}$H mass values obtained from production and from decay. The 
mass values obtained from production are systematically higher than those 
from decay by 
\begin{equation}
\Delta M({^{6}_{\Lambda}{\rm H}}) = 0.98 \pm 0.74~{\rm MeV}, 
\label{finaldeltamass} 
\end{equation} 
where the uncertainty is evaluated from the 1.3 MeV uncertainty for 
$T(\pi^{+})+T(\pi^{-})$ from which each of the mass differences is directly 
determined. Unlike the mean $M({^{6}_{\Lambda}{\rm H}})$ mass value, the 
spread of the production vs decay mass differences is well within $1\sigma$. 
A possible physical origin of the $\Delta M({^{6}_{\Lambda}{\rm H}})$ 
systematics is discussed in a subsequent section. 

The mean mass value (\ref{finalmass}) corresponds to a $^{6}_{\Lambda}$H 
binding energy $B_{\Lambda}=(4.0\pm1.1)$ MeV with respect to the 
$(\Lambda + {^{5}{\rm H}})$ threshold, and to $B_{\Lambda}=(0.3\pm1.1)$ MeV 
with respect to the lowest threshold $({_{\Lambda}^{4}{\rm H}}+2n)$. 
The $^{6}_{\Lambda}$H mean mass value and its uncertainty are indicated on 
the r.h.s. of Fig.~\ref{fig6} with respect to the various thresholds and 
predictions shown on the l.h.s. of the figure.

\section{Background estimation} 

Before discussing the physical interpretation of the above results, it is 
mandatory to check carefully that the three observed events do not arise 
from physical or instrumental backgrounds that could affect the data. 
Concerning the physical backgrounds, a complete simulation has been performed 
of possible $K^{-}_{\rm stop}$ absorption reactions on both single nucleons 
and pairs of strongly correlated nucleons that lead to the formation and decay 
of $\Lambda$ and $\Sigma$ hyperons. Of these reactions, only the following 
chain leads to ($\pi^{+},\pi^{-}$) coincidences in the same momentum ranges 
corresponding to the production and mesonic decay of $^{6}_{\Lambda}$H and 
which are respected by the three candidate events: 
\begin{eqnarray} 
K^{-}_{\rm stop}+{^{6}{\rm Li}} & \rightarrow & \Sigma^{+}+{^{4}{\rm He}}+n+
\pi^{-} \ \ \ \  (p_{\pi^{-}}\leq 190\ \mathrm{MeV/c}) \nonumber \\ 
& & \hookrightarrow n+\pi^{+} \ \ \ \ (p_{\pi^{+}}\leq 282\ \mathrm{MeV/c}). 
\label{bgd_S_1} 
\end{eqnarray} 

This reaction chain has been studied by means of the FINUDA simulation program 
fully reproducing the apparatus geometry, detection efficiency and the trigger 
efficiency. The interaction of $K^{-}$ with the target nucleus has been 
simulated with two different approaches. In the first approach, the quasi-free 
approximation was adopted for the interaction of the $K^{-}$ with a proton of 
the target nucleus, $K^{-}_{\rm stop}+p\to\Sigma^{+}+\pi^{-}$, taking into 
account the nucleon Fermi motion; the residual nucleus was considered as 
a spectator and the notation ``$^{4}$He + n" is just a label to indicate that 
the system is highly particle unstable. Pions arising from (\ref{bgd_S_1}) 
were processed by the pattern recognition and reconstruction programs of 
FINUDA as real data. In common with all simulated reaction chains, the 
simulated events were then submitted to the same quality cuts and to the same 
selections criteria applied in the data analysis. Three events were found out 
of a total of $2.2\cdot 10^{7}$ $K^{-}$ mesons simulated to stop on $^{6}$Li 
targets and forced to undergo the (\ref{bgd_S_1}) ``quasi-free" reaction chain 
with a probability of 1. Taking into account the number of actual $K^{-}$ 
mesons stopped on $^6$Li targets, the branching fraction for the 
$K^{-}_{\rm stop}+p\to\Sigma^{+}+\pi^{-}$ reaction on nuclei measured on 
$^{12}$C \cite{vander} and on $^{4}$He \cite{katz}, $(0.159\pm 0.012)$ 
evaluated as a weighted mean, the $\Sigma^{+}+n\rightarrow \Lambda+p$ 
conversion probability \cite{outa}, $(0.45\pm 0.04)$, and the 
$\Sigma^{+}\rightarrow n+\pi^{+}$ decay branching ratio, $(0.483\pm 0.003)$, 
an expected $\Sigma^{+}$ background of $0.15\pm 0.09$ events on $^{6}$Li 
targets is obtained. 

In a second approach, the interaction of $K^{-}$ mesons with the target 
nucleus as a whole was considered, applying directly the 4-body kinematics to 
(\ref{bgd_S_1}). Five events were found out of a total of $2.7\cdot 10^{7}$ 
$K^{-}$ mesons simulated to stop on $^{6}$Li targets and forced to undergo the 
(\ref{bgd_S_1}) ``4-body" reaction chain with a probability of 1. Taking into 
account the same normalization factors used for the ``quasi-free" approach, 
an expected $\Sigma^{+}$ reaction chain background of $0.20\pm 0.11$ events  
on $^{6}$Li targets was obtained under the hypothesis that $\Sigma^{+}$ 
production on $^{6}$Li in this approach always gives a recoiling $^{4}$He 
nucleus. Final states corresponding to further fragmentation of the $^{6}$Li 
target nucleus, such as $K^{-}_{\rm stop}+{^{6}{\rm Li}}\rightarrow\Sigma^{+}+
{^{3}{\rm He}}+n+n+\pi^{-}$, give weaker background contribution, owing to the 
requirements imposed on $T(\pi^{+})+T(\pi^{-})$ ($<180$ MeV for final states 
of the $\Sigma^{+}$ production reaction with more than 4 bodies) and on the 
$\pi^{+}$ and $\pi^{-}$ momenta. 

We also considered the distortion of Eq.~(\ref{bgd_S_1}) reaction chain 
spectra due to the $^{4}{\rm He}+n$ final state interaction leading to $^5$He,  
a resonance centered at $\sim 0.8$ MeV above the $^{4}{\rm He}+n$ threshold 
with $\Gamma=1.36$ MeV \cite{tilley02}. To this end we required that once the 
$^{4}$He and neutron momenta generated by the 4-body phase space simulation 
corresponded to the formation of the $^{5}$He resonance, the momenta of the 
remaining particles, $\Sigma^{+}$ and $\pi^{-}$, should be modified 
accordingly. We passed then these modified phase space distributions 
through the selection criteria described above and found variation of less 
than $1\%$ in the background value evaluated for a $100\%$ $^{4}{\rm He}+n$ 
final state. 

\begin{figure}[htbp] 
\begin{center} 
\includegraphics[width=90mm]{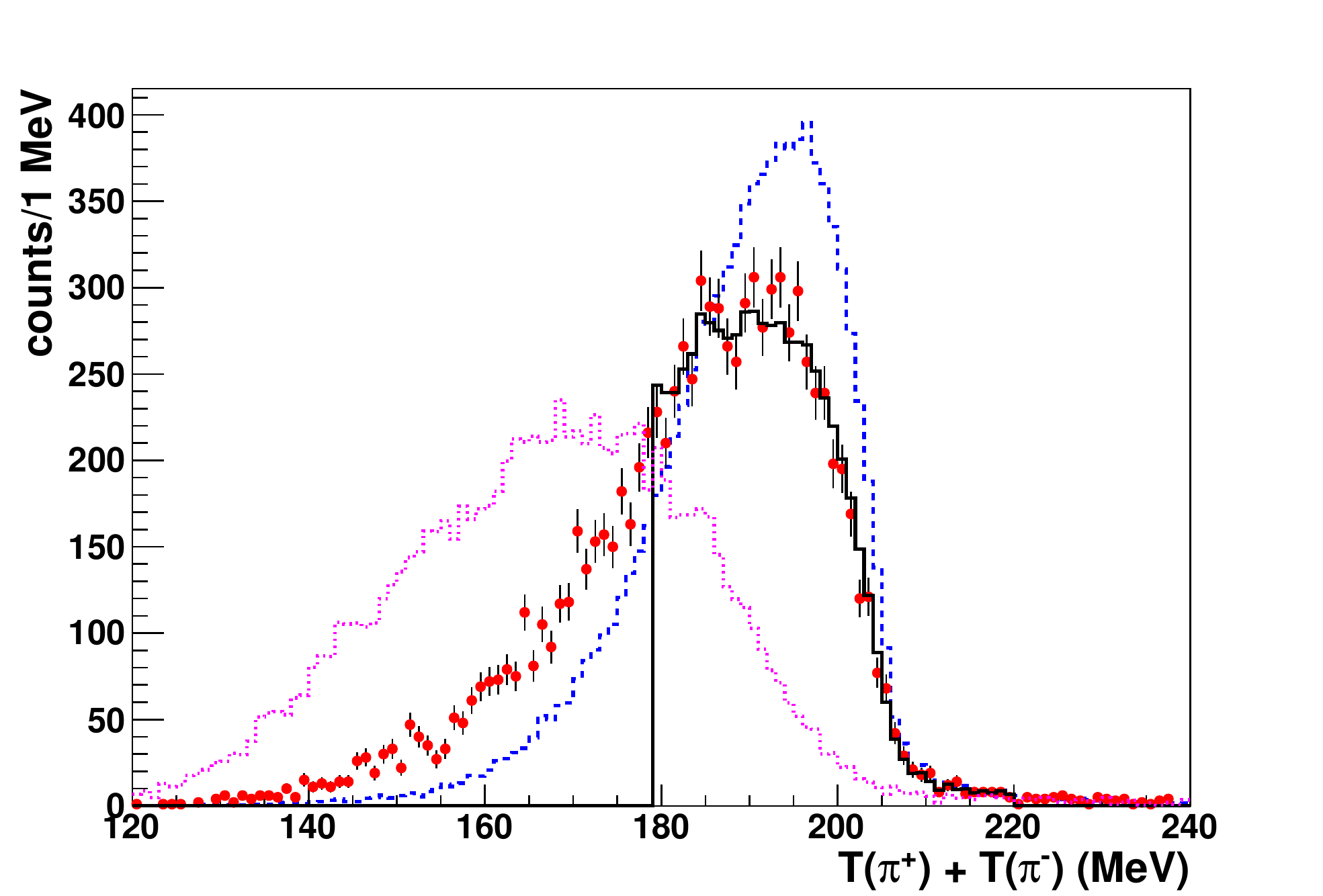} 
\caption{$T(\pi^{+})+T(\pi^{-})$ distributions; (red in the web version) points: experimental 
data; (blue in the web version) dashed histogram: ``quasi-free" simulation; (violet in the web version) dotted 
histogram: ``4-body" simulation; (black) continuous histogram: best fit to the 
data with fractions of the simulated templates. The simulated distributions 
have been normalized to the experimental distribution area. For more details 
see text.} 
\label{fig7} 
\end{center} 
\end{figure} 

In Fig.~\ref{fig7} the experimental $T(\pi^{+})+T(\pi^{-})$ spectrum is shown 
together with spectra obtained from the ``quasi-free" and ``4-body" 
simulations of the (\ref{bgd_S_1}) process: the simulated spectra were 
normalized to the area of the experimental distribution. As may be seen, the 
``quasi-free" spectrum (dashed (blue in the web version) histogram) reproduces the experimental 
distribution better than the ``4-body" (dotted (violet in the web version) histogram), but 
exhibits a too sharp decrease in the 200--210 MeV region and underestimates 
the low energy tail. To obtain a satisfactory description, a fit of the 
experimental spectrum was performed with fractions of the two simulated 
templates; a standard likelihood fitting method, using Poisson statistics, 
was applied in which both data and Monte Carlo statistical uncertainties were 
taken into account \cite{barlow}. Particular care was devoted to the 
description of the 200--210 MeV slope, where the selection of the 
$^{6}_{\Lambda}$H candidate events is made. The continuous (black in the web version) histogram 
in Fig.~\ref{fig7} represents the best fit to the 180--220 MeV region; 
the resulting fractions are $0.743\pm 0.019$ and $0.257\pm 0.017$ for the 
``quasi-free" and ``4-body" templates respectively, with a $\chi^{2}$/ndf = 
40.0/39. We note that varying the width of the fit region from 180--220 MeV 
to 130--220 MeV spoils the fit, increasing the $\chi^{2}$/ndf value by a 
factor of $\sim 3.8$, while the fractions of the two templates change by less 
than 0.025, corresponding to $1.3-1.5~\sigma$.  

Back in Fig.~\ref{fig4}, the dashed (red) histograms represent the separate 
$p_{\pi^{+}}$ and $p_{\pi^{-}}$ distributions obtained by adopting the above 
fractions to the events successfully simulated within the two approaches 
discussed above and satisfying the cut $T(\pi^{+})+T(\pi^{-})=202-204$ MeV. 
Only a qualitative agreement is reached with the experimental, low statistics 
distributions. The estimated background spectra are shifted from the 
experimental ones toward higher momentum for the $\pi^+$ spectrum and toward 
lower momentum for the $\pi^-$ spectrum. The difference is significant over 
the statistical fluctuation. However, these shifts cause overestimates of the 
background counts when the tails of the simulated distributions are used for 
an estimate. It is thus possible to conclude that the background estimate 
is safe in spite of the slight deviations noted here. In particular, in the 
$^{6}_{\Lambda}$H selected regions, indicated by shaded (blue in the web version) vertical bars, 
the contribution due to the $\Sigma^{+}$ background for events satisfying also 
the cuts on $p_{\pi^{+}}$ and $p_{\pi^{-}}$ in coincidence corresponds to 
$0.16\pm 0.07$ events on $^{6}$Li targets (BGD1). 

\vspace{3mm} 
Another reaction chain capable of providing background events is 
\begin{eqnarray} 
K^{-}_{\rm stop}+{^{6}{\rm Li}} & \rightarrow & {^{4}_{\Lambda}{\rm H}}+2n+
\pi^{+}\ \ \ \  (p_{\pi^{+}} \leq 252\ \mathrm{MeV/c}) \nonumber \\ 
& & \hookrightarrow {^{4}{\rm He}}+\pi^{-} \ \ \ \ (p_{\pi^{-}} \sim 132.8\ 
\mathrm{MeV/c}). 
\label{bgd_4LH_1} 
\end{eqnarray} 
The momentum of the $^{4}_{\Lambda}$H decay $\pi^{-}$ is close to the 
momentum of the $\pi^{-}$ from the two-body decay of $^{6}_{\Lambda}$H, 
$p_{\pi^{-}}\sim 134$ MeV/c, evaluated assuming $B_{\Lambda}=5$ MeV which is 
halfway between the two theoretical estimates exhibited in Fig.~\ref{fig6}. 

The probability of having background contribution from this reaction chain was 
evaluated taking into account the phase space fraction of the reaction 
(\ref{bgd_4LH_1}) available for $\pi^{+}$'s satisfying the momentum selection 
$p_{\pi^{+}}=250-255$ MeV/c, $4\cdot 10^{-6}$, and the probability for a 
$K^{-}_{\rm stop}$ to produce a $^{4}_{\Lambda}$H accompanied by a $\pi^{+}$ 
on a $^{6}$Li target. In Ref.~\cite{tamura_fr} the probability of producing 
$^{4}_{\Lambda}$H on $^{7}$Li targets was reported to be $(3.0\pm 0.4)\cdot 
10^{-2}/K^{-}_{\rm stop}$ and, furthermore, the probability of producing it 
together with a charged pion was indicated to be $0.49\pm 0.08$. Using these 
values, the formation probability of ${^{4}_{\Lambda}{\rm H}}+\pi^{\pm}$ 
on $^{6}$Li target, the closest isotope of $^{7}$Li, was evaluated to be 
$(1.47\pm 0.15)\cdot 10^{-2}/K^{-}_{\rm stop}$. In addition, a branching 
ratio 0.49 \cite{tamura_fr} for the two-body decay 
$^{4}_{\Lambda}{\rm H}\to {^{4}{\rm He}}+\pi^{-}$ has to be included. 
A total probability of $(2.8\pm 0.5)\cdot 10^{-8}$ is obtained. 
From this value, taking into account the global efficiency of FINUDA 
(acceptance, reconstruction and analysis cuts) it is possible to evaluate a 
background of $0.05\pm 0.01$ reconstructed events for the $2.7\cdot 10^{7}$ $K^{-}$ 
mesons stopped on the $^{6}$Li targets. It should be noted that this value 
overestimates by far the actual contribution from (\ref{bgd_4LH_1}) since the 
analysis of Ref.~\cite{tamura_fr} is incapable of separating the contribution 
of $^{4}_{\Lambda}{\rm H}+\pi^{+}$ out of the global $^{4}_{\Lambda}{\rm H}+
\pi^{\pm}$ fraction. Note that a microscopic reaction evaluation of 
(\ref{bgd_4LH_1}) requires a theoretical model which takes into account 
different channels ($\Lambda$ production, $\Sigma$ production, compound 
nucleus formation) the weights of which are not experimentally known. 
We chose to avoid relying on a model and to assume, instead, an overly 
conservative evaluation of the background from the chain (\ref{bgd_4LH_1}). 
The estimated level of this background is negligible with respect to the 
BGD1 background from the chain (\ref{bgd_S_1}). 

%\newpage 
%\vspace{3mm} 
Other reaction chains have been considered, such as: 
\begin{eqnarray} 
K^{-}_{\rm stop}+{^{6}{\rm Li}} & \rightarrow & 
\Sigma^{+}+{^{3}_{\Lambda}{\rm H}}+d+\pi^{-}\ \ \  (p_{\pi^{-}}\leq 165\ 
{\rm MeV/c}) \nonumber \\ 
 &   & \hookrightarrow n + \pi^{+} \ \ \  (p_{\pi^{+}} < 250\ {\rm MeV/c}) 
\label{bgd_m1} 
\end{eqnarray} 

\begin{eqnarray} 
K^{-}_{\rm stop}+{^{6}{\rm Li}} & \rightarrow & 
{^{3}_{\Lambda}{\rm H}}+3n+\pi^{+}\ \ \  (p_{\pi^{+}}\leq 242\ 
{\rm MeV/c}) \nonumber \\ 
& & \hookrightarrow {^{3}{\rm He}}+\pi^{-}\ \ \  (p_{\pi^{-}}\sim 115\ 
{\rm MeV/c}) 
\label{bgd_m2} 
\end{eqnarray} 

\begin{eqnarray} 
K^{-}_{\rm stop}+{^{6}{\rm Li}} & \rightarrow & 
\Lambda + {^{3}{\rm H}}+2n+\pi^{+}\ \ \  (p_{\pi^{+}}\leq 247\ 
{\rm MeV/c}) \nonumber \\ 
& & \hookrightarrow p+\pi^{-}\ \ \  ( p_{\pi^{-}} < 195\ {\rm MeV/c}) 
\label{bgd_m3} 
\end{eqnarray} 
These reaction chains may be safely discarded since all of them involve too 
low values of $T(\pi^{+})+T(\pi^{-})$, less than 190 MeV, which are way 
outside the cut applied on $T_{\rm tot}(\pi)$; the chain (\ref{bgd_m2}) may 
be discarded in addition by the cut imposed on $p_{\pi^{-}}$. 

Finally, another mechanism which could produce a ($\pi^{+},\pi^{-}$) pair 
in the final state is 
\begin{eqnarray} 
K^{-}_{\rm stop}+{^{6}{\rm Li}} & \rightarrow & 
{^{6}_{\Lambda}{\rm He}}+\pi^{0}\ \ \  (p_{\pi^{0}}\sim 280\ {\rm MeV/c}) 
\nonumber \\ 
& & \hookrightarrow {^{6}{\rm Li}}+\pi^{-}\ \ \  (p_{\pi^{-}} \sim 108\ 
{\rm MeV/c}), 
\label{bgd_2pi0} 
\end{eqnarray} 
followed by a reaction on another $^{6}$Li nucleus: 
\begin{equation} 
\pi^{0}+{^{6}{\rm Li}} \rightarrow {^{6}{\rm He}}+\pi^{+}\ \ \ 
(p_{\pi^{+}}\sim 280\ \rm{MeV/c\ in\ the\ forward\ direction}). 
\label{bgd_2pi0_3} 
\end{equation} 
However, the kinematics of (\ref{bgd_2pi0}) rules out a contribution from this 
reaction chain when applying the cuts on $\pi^{-}$ momenta. Moreover, the mean 
free path of a $\pi^{0}$ with momentum $\sim 280$ MeV/c, 
$l_{\rm free} < 0.1~\mu$m, strongly reduces the probability of the second 
reaction of the chain with respect to $\pi^{0}$ decay. 

\vspace{3mm} 
The main source of instrumental background could be the presence of fake 
tracks, due to fake signals from the detectors, misidentified as true events 
by the track reconstruction algorithms. For this purpose we analyzed the 
events with a $\pi^{+}$ and a $\pi^{-}$ emitted in coincidence with a $K^{-}$ 
stopping in a given nuclear target under the following criteria: \\ 
\begin{enumerate} 
\item events relative to target nuclei other than $^{6}$Li ($^{7}$Li, 
$^{9}$Be, $^{13}$C, $^{16}$O) were selected with the same selection criteria 
$T(\pi^{+})+T(\pi^{-})=202-204$ MeV, $p_{\pi^{+}}=250-255$ MeV/c and 
$p_{\pi^{-}}=130-137$ MeV/c, as for $^{6}$Li. Incidentally, for events coming 
from the (\ref{bgd_S_1}) reaction chain on nuclear targets heavier than 
$^{6}$Li, $T(\pi^{+})+T(\pi^{-})<202$ MeV by at least $2\sigma_{T}$ so that 
this criterion actually selects the instrumental background exclusively. 
Only one event was found, coming from $^{9}$Be target; \\ 
\item events relative to the $^{6}$Li targets were selected with values of 
$T(\pi^{+})+T(\pi^{-})=193-199$ MeV so as to search for neutron-rich 
hypernuclei produced on the other targets. No events were found. 
\end{enumerate} 

\vspace{2mm} 
Taking into account the number of $K^{-}_{\rm stop}$ detected in $^{6}$Li 
targets and in all the other targets, we concluded that $0.27\pm 0.27$ 
fake events should be expected from $^{6}$Li due to the instrumental 
background (BGD2). 

Combining together the expected number of events arising from physical and 
instrumental backgrounds that affect our selected data, $0.16\pm 0.07$ (BGD1) 
and $0.27\pm 0.27$ (BGD2), a total of $0.43\pm 0.28$ background events has 
been established. 
From this value, following Poisson statistics, we may state that the three 
observed $^{6}_{\Lambda}$H candidate events do not belong to the background 
distribution with a confidence level of $99\%$; the difference between the 
measured yield and the total expected background can thus be safely considered 
a $^{6}_{\Lambda}$H signal. 
The probability of observing three or more events from the background fluctuation 
following Poisson distribution with $\mu=0.43$ (BGD1+BGD2) or $\mu=0.16$ (only BGD1) is 
0.0096 or 0.0006, respectively. In terms of a {\it statistical significance} $S$ defined by 
$S=C/\sqrt{\rm BGD}$, with a signal $C=3-{\rm BGD}$, the statistical significance of the signal 
is 3.9 or 7.1, respectively.

\section{Production rate evaluation} 

Using the background estimates of the last section, it is possible to evaluate 
the product $R \cdot {\rm BR}(\pi^{-})$, where $R$ is the $^{6}_{\Lambda}$H 
production rate per stopped $K^{-}$ and ${\rm BR}(\pi^{-})$ is the branching 
ratio for the two-body weak decay 
$^{6}_{\Lambda}{\rm H}\rightarrow {^{6}{\rm He}}+\pi^{-}$: 
\begin{equation} 
R \cdot {\rm BR}(\pi^{-}) = \frac{3 - {\rm BGD1} - {\rm BGD2}}
{\epsilon(\pi^{+})\ \epsilon(\pi^{-})\ K^{-}_{\rm stop}({^{6}{\rm Li}})} = 
(1.3 \pm 0.9)\cdot 10^{-6} /K^{-}_{\rm stop}. 
\label{rate} 
\end{equation} 
In Eq.~(\ref{rate}), $\epsilon(\pi^{+})$ and $\epsilon(\pi^{-})$ indicate 
the global efficiencies for $\pi^{+}$ and $\pi^{-}$, respectively, including 
detection efficiency, geometrical and trigger acceptances and pattern 
recognition, reconstruction and selection efficiencies, all of which have 
been evaluated by means of the full FINUDA simulation code, well tested in 
calculations for other reactions in similar momentum ranges 
\cite{spectrFND,NPA775,mwd}. $K^{-}_{\rm stop}({^{6}{\rm Li}})$ is the number of 
$K^{-}$ detected at stop in $^{6}$Li targets. 

The value (\ref{rate}) has to be corrected for the purity of the $^{6}$Li 
targets used, 90$\%$, for the 0.77 $\sigma_{T}$ cut applied to $T(\pi^{+})+
T(\pi^{-})$, and for the fraction of $^{6}_{\Lambda}$H decaying in flight. 
In Ref.~\cite{tamura_fr} a contribution of $20\%$ is reported for the decay 
in flight of $^{4}_{\Lambda}$H produced on a $^{7}$Li target; extending this 
value also to $^{6}_{\Lambda}$H and considering that the cut applied to 
$p_{\pi^{-}}$, 130--137 MeV/c, allows to accept about one half of the pions 
emitted in flight, a correction factor of $10\%$ is evaluated. The corrected 
result is 
\begin{equation} 
[R \cdot {\rm BR}(\pi^{-})]_{\rm corr.}=(2.9 \pm 2.0)\cdot 
10^{-6}/K^{-}_{\rm stop}. 
\label{finalrate} 
\end{equation} 

By assuming ${\rm BR}(\pi^{-})=49\%$, in analogy to the weak decay 
$^{4}_{\Lambda}{\rm H}\rightarrow {^{4}{\rm He}}+\pi^{-}$ \cite{tamura_fr}, 
we find $R =(5.9 \pm 4.0)\cdot 10^{-6}/K^{-}_{\rm stop}$, fully consistent 
with the upper limit (\ref{Rpi+6}) obtained previously by FINUDA 
\cite{nrich1}. Although no theoretical calculation of this capture rate has 
been reported to provide direct comparison, the order of magnitude of the 
rate determined here is compatible with the interval of values calculated 
for production of heavier neutron-rich hypernuclei \cite{tretyak01} with 
stopped $K^{-}$ mesons and, as expected, is approximately three orders of 
magnitude lower than the capture rate for the production of ordinary 
particle-stable $\Lambda$ hypernuclei.

\section{Discussion} 

The binding energy deduced from the three measured events listed 
in Table~\ref{tab1} was recorded in Eq.~(\ref{finalmass}): 
$B_{\Lambda}({^{6}_{\Lambda}{\rm H}})=4.0\pm 1.1$ MeV with respect to 
$^{5}{\rm H}+\Lambda$, close to the value 
$B_{\Lambda}({_{\Lambda}^{6}{\rm He}})=4.18\pm 0.10$ MeV with respect to 
$^{5}{\rm He}+\Lambda$ for the other known $A=6$ hypernucleus \cite{juric}. 
It is in good accord with the estimate 4.2 MeV made originally by Dalitz 
and Levi Setti \cite{dalitz_setti} and confirmed by Majling \cite{majling}. 
It is lower by 1.8 MeV than the value 5.8 MeV suggested by Akaishi et 
al. \cite{akaishi}, leaving little room for an attractive contribution from 
a $\Lambda NN$ three-body force of a similar magnitude, 1.4 MeV, which in 
Akaishi's calculations arises from a coherent $\Lambda N-\Sigma N$ mixing 
model. This is consistent with a substantial weakening of $\Lambda N-\Sigma N$ 
mixing contributions for the excess $p$ shell neutrons in 
$_{\Lambda}^{6}{\rm H}$ with respect to the strong effect calculated in the 
$s$-shell hypernucleus $_{\Lambda}^{4}{\rm H}$ \cite{akaishi00}. Indeed, 
recent shell-model calculations by Millener indicate that $\Lambda N-\Sigma N$ 
mixing contributions to $B_{\Lambda}$ and to doublet spin splittings in the 
$p$ shell are rather small, about $(10\pm 5)\%$ of their contribution 
in $_{\Lambda}^{4}{\rm H}$ \cite{millener}. Nevertheless, given the 
measurement uncertainty of 1.1 MeV, one may not conclude that this 
$\Lambda NN$ force contribution is negligible, but only that its influence 
appears considerably lower than predicted. For illustration, see 
Fig.~\ref{fig6}. 

It is possible to avoid considering explicitly the $\Lambda N-\Sigma N$ 
mixing effect in the evaluation of $B_{\Lambda}({_{\Lambda}^{6}{\rm H}})$ 
by updating the shell-model (SM) argument used in Ref.~\cite{dalitz_setti}. 
We adopt a cluster model for $_{\Lambda}^{6}$H in terms of $_{\Lambda}^{4}$H 
plus two $p$-shell neutrons coupled to $J^{\pi}=0^{+}$ as in $^{6}$He g.s. 
The interaction of the $\Lambda$ hyperon with this dineutron cluster, 
including any $\Lambda nn$ force arising from $\Lambda N-\Sigma N$ mixing, 
may be deduced from $_{\Lambda}^{7}$He which consists of an $\alpha$ cluster 
plus precisely the same $\Lambda nn$ configuration under consideration in 
$_{\Lambda}^{6}$H. Subtracting $B_{\Lambda}({_{\Lambda}^{5}{\rm He}})=3.12\pm 
0.02$ MeV from $B_{\Lambda}({_{\Lambda}^{7}{\rm He}})$, with a value 
$B_{\Lambda}({_{\Lambda}^{7}{\rm He}})= 5.36\pm 0.09$ MeV obtained by 
extrapolating linearly from the known binding energies of the other members of 
the $A=7$ hypernuclear $T=1$ isotriplet (see Fig.~3, Ref.~\cite{hashimoto}), 
we obtain $2.24\pm 0.09$ MeV for the $\Lambda nn$ sum of two-body and 
three-body interactions involving the $\Lambda$ hyperon. The value of 
$B_{\Lambda}({_{\Lambda}^{6}{\rm H}})$ is then obtained adding this 2.24 MeV 
to $B_{\Lambda}({_{\Lambda}^{4}{\rm H}})=2.04\pm 0.04$ MeV \cite{juric}, 
so that $B_{\Lambda}^{\rm SM}({_{\Lambda}^{6}{\rm H}})=4.28\pm 0.10$ 
MeV.{\footnote{We thank Dr. D.J. Millener for alerting us to this estimate. 
A somewhat higher value, 
$B_{\Lambda}^{\rm SM}({_{\Lambda}^{6}{\rm H}})=4.60\pm 0.24$ MeV, is obtained 
if the preliminary value $B_{\Lambda}({_{\Lambda}^{7}{\rm He}})=5.68\pm 
0.03({\rm stat})\pm 0.22({\rm syst})$ MeV from the $(e,e'K^+)$ reaction in the 
JLab E01-011 experiment is used \cite{hashimoto}.}} We have thus recovered the 
estimate originally made by Dalitz and Levi Setti \cite{dalitz_setti}. 

As mentioned in the discussion of Table~\ref{tab1}, the $^{6}_{\Lambda}$H 
mass values obtained from production are systematically higher than 
the corresponding values obtained from decay, leading to a mass 
difference of $\Delta M({^{6}_{\Lambda}{\rm H}})=0.98\pm 0.74$ MeV, 
see Eq.~(\ref{finaldeltamass}). This suggests that $^{6}_{\Lambda}$H 
is produced in an excited state, while decaying from its ground state. 
We recall that Pauli spin is conserved in capture at rest. 
For $K^{-}_{\rm stop}+{^{6}{\rm Li}}\to {^{6}_{\Lambda}{\rm H}}+\pi^{+}$ 
production, since $^{6}$Li is very well approximated (about $98\%$) by a 
$L=0,~S=1$ configuration \cite{millener}, $^{6}_{\Lambda}$H is dominantly 
produced in its $1^{+}$ first excited state, decaying then by a fast magnetic 
dipole transition to the $0^{+}$ ground state from which the mesonic weak 
decay occurs. In this situation, the pion kinetic energies $T(\pi^{+})$ and 
$T(\pi^{-})$, directly measured by the FINUDA spectrometer, should reflect 
this systematic difference between production and decay. The mass of the 
$0^{+}$ ground state should be calculated from the decay reaction only, 
giving a mean value $M({^{6}_{\Lambda}{\rm H}_{\rm g.s.}})=(5800.9\pm 1.2)$ 
MeV, corresponding to a binding energy of $4.5\pm 1.2$ MeV with respect 
to ($\Lambda+{^{5}{\rm H}}$) and of $0.8\pm 1.2$ MeV with respect to 
(${^{4}_{\Lambda}{\rm H}}+2n$). For the $1^{+}$ excited state it should be 
possible to evaluate a mean mass $M({^{6}_{\Lambda}{\rm H}^{\ast}})=(5801.9\pm 
1.0)$ MeV. Although nominally unstable by $0.2\pm 1.0$ MeV, the low $Q$ value 
for two-neutron emission plus the associated $\Delta S=1$ spin flip required 
in the decay to $^{4}_{\Lambda}{\rm H}(0^{+}_{\rm g.s.})+2n$ are likely to 
make the $^{6}_{\Lambda}{\rm H}(1^{+}_{\rm exc}\to 0^{+}_{\rm g.s.})$ $M1$ 
$\gamma$-ray transition competitive with the strong decay of the 
$1^{+}_{\rm exc}$ level.

It is worth noting that the uncertainty placed on the excitation 
energy $\Delta E (0^{+}\to 1^{+})$, identifying this $\Delta E$ with 
$\Delta M({^{6}_{\Lambda}{\rm H}})$, is considerably smaller than 
the uncertainty of each one of the $0^{+}$ and the $1^{+}$ levels because 
$\Delta M$ has been determined directly from the sum of kinetic energies 
$T(\pi^{+})+T(\pi^{-})$ and its associated uncertainty. The value determined 
in the present experiment for $\Delta E$ is smaller by $2\sigma$ than 
the value $\Delta E_{\rm akaishi}(0^{+}\to 1^{+})=2.4$ MeV predicted in 
Ref.~\cite{akaishi}. This is in line with the conclusion drawn from the 
absolute energy location of the $0^{+}$ g.s., casting doubts on the 
applicability of the model developed by Akaishi et al.~\cite{akaishi}. The 
value $\Delta E({^{6}_{\Lambda}{\rm H}}:0^{+}\to 1^{+})=0.98\pm 0.74$ MeV is 
in good accord with $\Delta E({^{4}_{\Lambda}{\rm H}}:0^{+}\to 1^{+})=1.04\pm 
0.03$ MeV \cite{juric}, consistently with a weak-coupling picture for the two 
`halo' $p$-shell neutrons in $^{6}_{\Lambda}{\rm H}$ outside the $s$-shell 
cluster of $^{4}_{\Lambda}{\rm H}$. 

It is also worth noting that the width of the selected $T(\pi^{+})+T(\pi^{-})$ 
interval (2 MeV) allows to include both production and decay pions within the 
experimental resolution of the mass determination at a $1\sigma$ level, thus 
extending the validity of the working assumption on which the analysis method 
was based, namely that the masses of the produced and decaying hypernucleus 
are equal. A variation of the binding energy from the value of 5 MeV used to 
fix the selection on $T(\pi^{+})+T(\pi^{-})$ to 4 MeV, the average binding 
energy of the $0^+$ and $1^+$ levels, produces a completely negligible 
variation of the accepted fraction of events due to the selection criteria 
with respect to the errors. Finally, the absence of systematics arising 
from the particular choice of both width and position of the selected 
$T(\pi^{+})+T(\pi^{-})$ interval indicates that the difference between the 
$^{6}_{\Lambda}$H formation and decay masses is not influenced by the cut 
itself. It is important, however, to realize that the above deductions on 
the $^{6}_{\Lambda}$H excitation spectrum rely on very scarce statistics and, 
therefore, have to be considered as indication, even if quite solid. 

Before closing we wish to discuss briefly another scenario for the excitation 
spectrum of $^{6}_{\Lambda}{\rm H}$ motivated by the somewhat large spread 
among the three $^{6}_{\Lambda}{\rm H}$ candidate events. Apart from the $0^+$ 
g.s. and $1^+$ spin-flip excited state as in $_{\Lambda}^{4}{\rm H}$, a $2^+$ 
excited state as for the $p$-shell dineutron system in $^{6}{\rm He}$ (1.80 
MeV) is expected at about 2 MeV excitation in $^{6}_{\Lambda}{\rm H}$. 
Furthermore, a triplet of spin-flip excitations $1^+,2^+,3^+,$ built on the 
$2^+$ dineutron excitation is expected 1 MeV higher, at about 3 MeV excitation 
in $^{6}_{\Lambda}{\rm H}$. It is then not unreasonable to assign event 3 in 
Table~\ref{tab1} to formation and decay of $^{6}_{\Lambda}{\rm H}$ involving 
the $0^+$ g.s. and its $1^+$ spin-flip excited state, as considered above, 
whereas the other two events which are relatively close to each other 
correspond to formation of one of the $1^+,2^+,3^+$ levels and to decay 
from the $2^+$ dineutron excitation. This scenario generates an additional 
excitation scale to confront the $\approx 3$ MeV separation between the first 
two events of Table~\ref{tab1} and the third one. This results in the 
assignment of $^{6}_{\Lambda}{\rm H}$ levels listed in Table~\ref{tab2}. 

\begin{table}[h] 
\begin{center} 
\caption{Masses and $B_{\Lambda}$ values (in MeV) of $^{6}_{\Lambda}$H levels 
assuming that event 3 in Table~\ref{tab1} corresponds to the lowest two 
levels and events 1 and 2 correspond to higher levels.} 
\label{tab2} 
\vspace{2mm} 
\begin{tabular}{ccccc} 
\hline 
$^{6}_{\Lambda}$H & $0^+$ & $1^+$ & $2^+$ & $(1^+,2^+,3^+)$ \\ 
\hline 
$M$ \ & \ 5798.66$\pm$0.84 \ & \ 5799.97$\pm$0.96 \ & \ 5802.07$\pm$0.59 \ & 
\ 5802.89$\pm$0.68 \\ 
$B_{\Lambda}$ \ & \ 6.78$\pm$0.84 \ & \ 5.47$\pm$0.96 \ & \ 3.37$\pm$0.59 \ & 
\ 2.55$\pm$0.68 \\
\hline 
\end{tabular} 
\end{center} 
\end{table} 

The table exhibits that the g.s. of $^{6}_{\Lambda}$H is bound in this 
scenario much stronger than the SM estimate outlined above, and in fact it 
is even more bound than predicted by Akaishi \cite{akaishi}, although the 
excitation energy of the spin-flip $1^+$ level appears considerably smaller 
than in his prediction. The excitation energy of the $2^+$ level comes 
out about 3.4 MeV, too high with respect to the simple SM consideration. 
We conclude that this scenario is unlikely, but this conclusion does not 
derive from any model-independent experimental observation. 
Future experiments will tell. 

\section {Conclusions} 

We have reported the first observation of the hyper superheavy hydrogen 
$^{6}_{\Lambda}$H, based on detecting 3 candidate events that cannot be 
attributed to pure instrumental or physical backgrounds. The resulting binding 
energy of $^{6}_{\Lambda}$H, $B_{\Lambda}=4.0\pm 1.1$ MeV, agrees with simple 
shell-model estimates initiated by Dalitz and Levi Setti \cite{dalitz_setti}, 
but disagrees with the prediction made by Akaishi \cite{akaishi} based on 
a strongly attractive $\Lambda NN$ interaction within a coherent 
$\Lambda N - \Sigma N$ mixing model. It was suggested that the excitation 
energy of the $1^+$ spin-flip state with respect to the $0^+$ g.s. be 
identified with the systematic difference $\Delta M=0.98\pm 0.74$ MeV between 
values of $^{6}_{\Lambda}$H mass derived separately from production and from 
decay. This value is consistent with the 1.04 MeV for the analogous spin-flip 
excitation in $^{4}_{\Lambda}$H, confirming again the applicability of the 
shell-model estimates. An experiment to produce $^{6}_{\Lambda}$H via the 
($\pi^{-}, K^{+}$) reaction on $^{6}$Li at 1.2 GeV/c was recently approved 
at J-PARC \cite{P10} and should run soon. The expected energy resolution is 
2.5 MeV FWHM, and the expected statistics about 1--2 orders of magnitude 
higher than previous KEK experiments.

%\vspace{5mm} 
\section*{Aknowledgements} 

We dedicate this article to the memory of our colleague Ambrogio Pantaleo, 
prematurely passed away. Dr. Pantaleo participated actively since the 
beginning to the FINUDA experiment and started the study of neutron-rich 
hypernuclei production in FINUDA, first results of which have been published 
in \cite{nrich1}; this article marks the completion of his dedicated 
contribution.

\end{document}